\documentclass{article}
\usepackage{graphicx}
\usepackage{amsmath}
\usepackage{amsopn}
\usepackage{amsfonts}
\usepackage{amssymb}
\usepackage[square]{natbib}
\usepackage{url}
\usepackage{subfigure}
\usepackage{booktabs}
\usepackage{rotating}
\usepackage[hang,footnotesize,bf]{caption}
\usepackage{setspace}
\usepackage{multirow}
\usepackage[]{algorithm}
\usepackage{algorithmic}
\usepackage{epstopdf}

\usepackage{colortbl}
\arrayrulecolor{black}
\doublerulesepcolor{black}

\newcommand*{\IsDraft}{false}

\newcommand*{\mediumwidth}{10cm}
\newcommand*{\halfwidth}{8cm}

\newcommand{\be}{\begin{equation}}
\newcommand{\ee}{\end{equation}}
\newcommand{\ba}{\begin{eqnarray}}
\newcommand{\ea}{\end{eqnarray}}

\begin{document}
\title{Earthquake Forecasting Based on Data Assimilation: \\
Sequential Monte Carlo Methods for Renewal Processes}

\author{
Maximilian J.\ Werner,\textsuperscript{1}
Kayo Ide,\textsuperscript{2}
Didier Sornette,\textsuperscript{3}}
\date{\today}
\maketitle
\textsuperscript{1} Swiss Seismological Service, Institute of Geophysics, ETH Zurich, Switzerland. (mwerner@sed.ethz.ch)

\textsuperscript{2} Department of Atmospheric and Oceanic Science,
Center for Scientific Computation and Mathematical Modeling,
Institute for Physical and Scientific Technology, 
University of Maryland, College Park, USA.

\textsuperscript{3} Department of Management, Technology and Economics, 
and Department of Earth Sciences, ETH Zurich, Switzerland.

\doublespacing

\section*{Abstract}
In meteorology, engineering and computer sciences, data assimilation is
routinely employed as the optimal way to combine noisy observations with
prior model information for obtaining better estimates of a state, and thus
better forecasts, than can be achieved by ignoring data
uncertainties. Earthquake forecasting, too, suffers from  measurement
errors and partial model information and may thus gain significantly
from data assimilation. We present perhaps the first fully implementable
data assimilation method for earthquake forecasts generated by a
point-process model of seismicity. We test the method on a synthetic and
pedagogical example of a renewal process observed in noise, which is
relevant to the seismic gap hypothesis, models of characteristic
earthquakes and to recurrence statistics of large quakes inferred from
paleoseismic data records. To address the non-Gaussian statistics of
earthquakes, we use sequential Monte Carlo methods, a set of flexible
simulation-based methods for recursively 
estimating arbitrary posterior distributions. We perform extensive
numerical simulations to demonstrate the feasibility and benefits of
forecasting earthquakes based on data assimilation. In particular, we
show that the forecasts based on the Optimal Sampling Importance
Resampling (OSIR) particle filter are significantly better than those of
a benchmark forecast that ignores uncertainties in the observed event
times. We use the marginal data likelihood, a measure of the explanatory
power of a model in the presence of data errors, to estimate parameters
and compare models.

\tableofcontents

\section{Introduction}

In dynamical meteorology, the 
primary purpose of data assimilation has been 
to estimate and forecast as accurately as possible the state of
atmospheric flow, using 
all available appropriate information \citep{Talagrand1997}. 
Recent advanced methods of data assimilation attempt to  include the effects of uncertainties explicitly in the estimation by taking probabilistic approaches.
\citet{Kalnay2003} defines data assimilation as a
statistical combination of observations and short-range
forecasts. According to \citet{WikleBerliner2007}, data assimilation is
an approach for fusing data (observations) with prior knowledge (e.g.,
mathematical representations of physical laws or model output) to obtain
an estimate of the distribution of the true state of a process. To
perform data assimilation, 
three components are required: (i) a 
statistical model for observations (i.e., a data or measurement model)
and (ii) an a priori statistical model for the state process (i.e., a
state or process model), which may be obtained through a physical
model of the time-evolving system, and (iii) a method to effectively 
merge the information from (i) and (ii). 

Both data and model are affected by uncertainty, due to measurement and
model errors and/or stochastic model elements, leading to uncertain
state estimates that can be described by probability distributions. Data
assimilation is therefore a Bayesian estimation problem: the prior is
given by model output (a forecast from the past) and the likelihood by the
measurement error distribution of the data. The posterior provides the best estimate
of the true state and serves as initial condition for a new
forecast. The essence of data assimilation is to inform uncertain data
through the model, or, equivalently, to correct the model using the
data. The cycle of predicting the next state and updating, or correcting
this forecast given the next observation, constitutes sequential data
assimilation (see \citep{Daley1991, Ghil1991, Ide-et-al1997,
Talagrand1997, Kalnay2003} for introductions to data assimilation and
\citep{Tarantola1987, Miller-et-al1999, Pham2001, WikleBerliner2007} for
a Bayesian perspective). 

Although data assimilation is increasingly popular in meteorology, climatology, oceanography, computer sciences, engineering and finance, only a few partial attempts, reviewed in section \ref{sec:lit}, have been made within the statistical seismology community to use the concept for seismic and fault activity forecasts. But earthquake forecasting suffers from the same issues encountered in other areas of forecasting: measurement uncertainties in the observed data and incomplete, partial prior information from model forecasts. Thus, basing earthquake forecasting on data assimilation may provide significant benefits, some of which we discuss in section \ref{sec:why}. Indeed, there is a growing perception in the seismological and geophysical community
that data assimilation should be used in earthquake forecasts. For instance,
at the inaugural meeting of the
ACES international cooperation on earthquake simulations (\url{http://www.aces.org.au/}), it
was concluded that there 
is a strong need for a gradual evolution from the well-developed concept of geophysical
data inversion to the emerging one of data assimilation, for which no generally accepted method exists in the domain of earthquake modeling. 

There are perhaps two major challenges for developing data assimilation
methods for earthquake forecasts: Firstly, seismicity is often modeled
by point-processes, and secondly, earthquake statistics are far from
Gaussian. Neither are typical assumptions in practical data assimilation
methods. To explain the first issue, data assimilation is often cast in
terms of discrete-time state-space models, or Hidden Markov models
(HMMs), reflecting the underlying physics-based stochastic differential
equations \citep{Daley1991, Kalnay2003, Kuensch2001, Cappe-et-al2005,
Doucet-et-al2001}.  An HMM is, loosely speaking, a Markov chain observed
in noise \citep{Doucet-et-al2001, DurbinKoopman2001, Kuensch2001, 
RobertCasella2004, Cappe-et-al2005, Cappe-et-al2007}: An HMM consists of an
unobserved Markov (state) process and an associated, conditionally
independent observation process (both process being potentially
nonlinear/non-Gaussian; see section \ref{sec:DAHMM} for precise definitions). The Kalman filter is an archetypical assimilation method for such a model \citep{Kalman1960, KalmanBucy1961}. In contrast, 
earthquake catalogs have many features which make them
uniquely distinct from the forecast targets in other disciplines. Indeed, seismicity models are usually stochastic point processes \citep{DaleyVereJones2003, VereJones1970, VereJones1995, KaganKnopoff1987, Ogata1998, HelmstetterSornette2002, RhoadesEvison2004, KaganJackson2000}, which
are completely different from the noisy differential or finite difference equations decorated by noise 
of standard data assimilation methods. There seems to exist little statistical work that extends the idea of data assimilation or state filtering to point processes, which model the stochastic point-wise space-time occurrence of events along with their marks. 

The second challenge, that of non-Gaussian probability distributions, has been solved to some extent by recent Monte Carlo methods, especially for models with a small number of dimensions \citep{Liu2001,
Doucet-et-al2001, RobertCasella2004}. In particular, Sequential Monte
Carlo (SMC) methods, a set of simulation-based methods for recursively
estimating arbitrary posterior distributions, provide a flexible,
convenient and (relatively) computationally-inexpensive method for
assimilating non-Gaussian data distributions into nonlinear/non-Gaussian
models \citep{Doucet-et-al2001, DurbinKoopman2001, Kuensch2001,
RobertCasella2004, Cappe-et-al2005, Cappe-et-al2007}. Also called particle filters, SMC filters have been
particularly successful at low-dimensional filtering problems for the
family of HMMs or state-space
models.
The Kalman-L\'{e}vy filter \citep{SornetteIde2001} provides an analytic solution extending
the Kalman filter for L\'{e}vy-law 
and power-law distributed model errors and data uncertainties.
We present an overview of SMC methods in sections \ref{sec:SMC} and \ref{sec:PF}.

The main purpose of this article is to develop an implementable method for forecasting earthquakes based on data assimilation. We test this sequential method on a pedagogical and synthetic example of a simulated catalog of ``observed" occurrence times of earthquakes, which are not the ``true" event times because of observational errors. We specifically use a point-process as our model of seismicity. To estimate arbitrary posterior distributions of the ``true" event times, we use the SMC methods we just mentioned. 

Our technique offers a step towards the goal of developing a ``brick-by-brick" approach to earthquake predictability \citep{Jordan2006, Jackson1996, Kagan1999}, given the enormous difficulties in identifying reliable 
precursors to impending large earthquakes \citep{Geller1997, Geller-et-al1997a, Kagan1997b}. 
With suitable adaptations and extensions, our approach should find its natural habitat in the general 
testing framework developed within the
Regional Earthquake Likelihood Models (RELM) Working Group \citep{Field2007, Schorlemmer-et-al2007, Schorlemmer-et-al2009} and the international Collaboratory for the Study of Earthquake Predictability (CSEP) \citep{Jordan2006, Zechar-et-al2009}, in which 
forecast-generating models are tested in a transparent, controlled, reproducible and fully prospective manner. 

The importance of  data uncertainties, forecast specification and evaluation for earthquake predictability experiments were highlighted by several recent studies. \citet{WernerSornette2008a} showed that measurement errors in magnitudes have serious, adverse effects on short-term forecasts performed using a general class of models of clustered seismicity, including two of the most popular models, the Short Term Earthquake Probabilities (STEP) model \citep{Gerstenberger-et-al2005} and the Epidemic-Type Aftershock Sequence (ETAS) model \citep{Ogata1988}. Moreover,   \citet{WernerSornette2008a} showed that the RELM evaluation tests are not appropriate for the broadened forecast distributions that arise from taking into account uncertainties in data, and recommended that forecasts should be replaced by a full distribution.  \citet{Schorlemmer-et-al2009} confirmed and supported this recommendation after examining first results from the five-year (i.e. long-term) RELM forecast competition. The methods used in this article for evaluating point-process forecasts when the observations are noisy provide an alternative to the current forecast evaluation method used in RELM and CSEP. 

Data and parameter uncertainties also play a crucial role in the ongoing debate about the relevance of the seismic gap hypothesis \citep{McCann-et-al1979, Nishenko1991, KaganJackson1991a, KaganJackson1995, Rong-et-al2003, McGuire2008}, of models of characteristic earthquakes \citep{Wesnousky1994, Bakun-et-al2005, Scholz2002, Kagan1993}, and of recurrence statistics of earthquakes on a particular fault segment inferred from paleoseismic data records \citep{Bakun-et-al2005, Davis-et-al1989, Rhoades-et-al1994, Ogata1999, Ogata2002, SykesMenke2006, Parsons2008}. The data are often modeled using renewal processes, and studies investigating data and parameter uncertainty confirmed that any model inference or forecast must take into account uncertainties \citep{Davis-et-al1989, Rhoades-et-al1994, Ogata1999, Ogata2002, SykesMenke2006, Parsons2008}. 

In this article, we focus on the class of renewal processes as models of seismicity. On the one hand, renewal processes are extensively used to model paleoseismic data records, characteristic earthquakes and seismic gaps, as mentioned above. On the other hand, renewal processes are the point-process analog of Markov chains, thereby enabling us to use sequential Monte Carlo methods developed for state-space models. In other words, renewal processes are the simplest class of point process models relevant to statistical seismology. By developing rigorously a data assimilation
procedure for renewal processes, we aim at providing the building blocks for more complicated models. In addition to the obvious relevance to earthquake forecasts, we hope to generate interest among statisticians to tackle the general problem of state filtering for point processes, for which the Markovian state-space model framework seems too restrictive. 

The article is structured as follows. Section \ref{sec:bg} provides a
brief literature review of data assimilation in connection with
statistical seismology and points out potential benefits of data
assimilation to earthquake forecasting. Section \ref{sec:method}
introduces the methods we believe are relevant in the seismicity
context, 
Section \ref{sec:DAHMM} provides the notation and basic Bayesian
estimation problem we propose to solve for renewal processes.
Section \ref{sec:Renew} defines renewal processes, which serve as our forecast models.
Section \ref{sec:SMC} explains the basics of Sequential Monte Carlo methods. In section \ref{sec:PF}, we describe three particular SMC filters (often called particle filters) in order of sophistication. To perform model inference, we must estimate parameters, which is described in section \ref{sec:PE}. Section \ref{sec:NumExp} describes numerical experiments to demonstrate how earthquake forecasting based on data assimilation can be implemented for a particular renewal process, where inter-event times are lognormally distributed. Section \ref{sec:NumExpSetUp} describes the set-up of the simulations: we use a lognormal renewal process of which only noisy occurrence times can be observed. In section \ref{sec:NumExpFiltering} we use particle filters to estimate the actual occurrence times, demonstrating that the particle filters improve substantially on a forecasting method that ignores the presence of data uncertainties. In section \ref{sec:NumExpPE}, we show that parameter estimation via maximum (marginal) likelihood is feasible. We conclude in section \ref{sec:Con}.

\section{Data Assimilation and Probabilistic Earthquake Forecasting}
\label{sec:bg}
\subsection{Literature on Probabilistic Earthquake Forecasting and Data Assimilation}
\label{sec:lit}

The general concepts of data assimilation or Hidden Markov models (HMMs) state inference are
relatively new to earthquake seismology and statistical earthquake
modeling. The few studies that are related can be separated into three
categories. (i)  \citet{Varini2005, Varini2008} studied a HMM of seismicity, in which the (unobserved) state could be in one of three different states (a Poisson process state, an ETAS process state and a stress-release process state) and the observational data were modeled according to one of the three processes. Varini did not consider measurement uncertainties of the data. 
(ii) \citet{GrantGould2004} proposed data formats and standards for the assimilation of uncertain paleoseismic data into earthquake simulators. \citet{VanAalsburg-et-al2007} assimilated uncertain paleoseismic data into ``Virtual California", a fixed-geometry earthquake simulator of large earthquakes: model runs are accepted or rejected depending on whether simulated earthquakes agree with the paleoseismic record. 
(iii) \citet{Rhoades-et-al1994} calculated seismic hazard on single fault segments by averaging the hazard function of a renewal process over parameter and data uncertainties, achieved by sampling over many parameter and data samples.  \citet{Ogata1999} presented a Bayesian approach to parameter and model inference on uncertain paleoseismic records, closely related to our approach. Data uncertainties were represented with either a uniform or a triangular distribution. To compute the integrals, Ogata seems to have used numerical integration, a process that becomes increasingly difficult as the number of events increases, in contrast to the particle filters that we use below. \citet{SykesMenke2006} assumed Gaussian data errors and uncorrelated recurrence intervals, also providing a maximum likelihood estimation procedure for the parameters of a lognormal process based on a Monte Carlo integration approach. \citet{Parsons2008} provided a simple but inefficient Monte Carlo method for estimating parameters of renewal processes from paleoseismic catalogs. 

\subsection{Why base earthquake forecasting on data assimilation?}
\label{sec:why}
The brief review of the literature indicates that the potential of data assimilation for earthquake forecasting has not been exploited. Some of the potential benefits are summarized here:

\begin{itemize}

\item Earthquake forecasting under observational uncertainties: The current surge in earthquake predictability experiments \citep{Field2007, Jordan2006, Schorlemmer-et-al2007, Schorlemmer-et-al2009, Zechar-et-al2009} provides strong motivational grounds for developing earthquake forecasting methods that are robust with respect to observational uncertainties in earthquake catalogs. 

\item Seismic hazard calculations under observational uncertainties: On a practical level, the inclusion of uncertainties for seismic hazard calculations may provide better scientific support for decision-making for the insurance industry, disaster agencies and risk mitigation strategies. 

\item Scientific hypothesis testing under observational uncertainties: Observational uncertainties may bias conclusions regarding scientific hypotheses. The only way to accurately provide confidence limits to test our ideas about earthquakes is to take uncertainties into account -- data assimilation provides the ideal vehicle for such a systematic treatment. 

\item Model inference: Data assimilation can be used as a framework for likelihood-based model testing and development, fully accounting for uncertainties. 

\item Near-Real Time Forecasting: Once new observations become available, data assimilation provides a vehicle for correcting an existing forecast without having to re-calibrate and re-initialize the model on the entire data set. For complex, high-dimensional models, this reduced computational burden may allow near-real-time forecasts.

\item Estimating physical quantities of physics-based models from seismicity: In its general formulation as a state and parameter estimation problem, data assimilation may also be viewed as a method for estimating physical quantities (``states") and model parameters, directly related to physics-based models, such as rate-and-state friction and Coulomb stress-change models. 

\item Optimally integrating different types of data for earthquake forecasts: The models submitted to the five-year forecast competition of RELM contain a variety of data from which forecasts are generated. In the future, the coupled integration of several types of different data is highly desirable. Numerical weather prediction has a long history of integrating such different types of data -- statistical seismology may be able to adapt these methods. 

\item Developing statistical theory and methodology for inference of point processes under observational data uncertainties: The theory of point processes has so far largely focused on exact data. The development of the statistical theory and practical methodology for taking into account noisy observations is therefore interesting for applications beyond earthquake forecasting. 

\end{itemize}

\section{Method: Sequential Monte Carlo Methods for Renewal Processes}
\label{sec:method}
\subsection{Bayesian Data Assimilation of State-Space or Hidden Markov Models (HMMs)}
\label{sec:DAHMM}

In this section, we state the general problem of Bayesian data assimilation that will be solved for specific model and observation assumptions in section \ref{sec:NumExp}. The presentation borrows from \citep{Doucet-et-al2000, Doucet-et-al2001} and \citep{Arulampalam-et-al2002}; see also \citep{Kuensch2001, RobertCasella2004, Cappe-et-al2005, Cappe-et-al2007, WikleBerliner2007} and references therein. 

We restrict ourselves to signals modeled as Hidden Markov Models (HMMs), i.e. Markovian, nonlinear, non-Gaussian state-space models. The unobserved signal (the hidden states) $\{x_t \}_{t \geq 1}$ is modeled as a Markov process ($x_t$ may be a vector). The initial state $x_0$ has initial distribution $p(x_0)$. The transition from $x_t$ to $x_{t+1}$ is governed by a Markov transition probability distribution $p(x_{t+1}|x_t)$. The observations $\{y_t \}_{t \geq 1}$ are assumed to be conditionally independent given the process $\{x_t \}_{t \geq 1}$ and of conditional distribution $p(y_t|x_t)$ (the observations may also be vectors, in general of different dimension than the state). The model can be summarized by
\ba
\text{Initial condition:} &p(x_0) & \label{eq:IC} \\
\text{Model forecast: } &p(x_{t+1}|x_t) & \hspace{1cm} t \geq 0 \label{eq:model} \\
\text{Conditional data likelihood: } & p(y_t|x_t) & \hspace{1cm} t \geq 1 \label{eq:condLL} 
\label{HMM}
\ea

We denote $x_{0:t} = \{x_0, \dots, x_t \}$ and $y_{1:t} = \{y_1, \dots, y_t \}$. The problem statement is then as follows: the aim is to estimate sequentially in time the posterior distribution $p(x_{0:t}| y_{1:t})$. We may also be interested in estimating the marginal distribution $p(x_{t}| y_{1:t})$, also known as the filtering distribution, and the marginal complete data likelihood $p(y_{1:t})$, which we will use for parameter estimation.

At any time $t$, the posterior distribution is given by Bayes' theorem
\be
\text{Bayes' Theorem:} \hskip 0.5cm p(x_{0:t}|y_{1:t}) =  \frac{p(y_{1:t}|x_{0:t}) \ p(x_{0:t})  }{\int p(y_{1:t}|x_{0:t}) \ p(x_{0:t}) dx_{0:t}}
\label{bayesDA}
\ee

A recursive or sequential formula can be derived from (i) the Markov property of the state process and (ii) the independence of observations given the state:  
\be
\text{Sequential Bayes' Theorem:} \hskip 0.5cm p(x_{0:t+1}|y_{1:t+1})=p(x_{0:t}|y_{1:t}) \ \frac{p(y_{t+1}|x_{t+1}) \ p(x_{t+1}|x_t)}{p(y_{t+1}|y_{1:t})}
\label{bayesrec}
\ee
where $p(y_{t+1}|y_{1:t})$ is given by
\be
p(y_{t+1}|y_{1:t}) = \int \dots \int p(y_{t+1}|x_{t+1}) p(x_{t+1}|x_t) p(x_{0:t}|y_{1:t}) dx_{0:t}
\label{eq:seqLL}
\ee

The marginal distribution $p(x_t|y_{1:t-1})$ also satisfies the
following recursion: 
\ba
&\text{Prediction:} &\ p(x_t|y_{1:t-1})= \int p(x_t|x_{t-1}) p(x_{t-1}|y_{1:t-1}) dx_{t-1} \label{predict} \\
&\text{Updating:} &\ p(x_t|y_{1:t}) = \frac{p(y_t|x_t) \ p(x_t|y_{1:t-1})}{\int p(y_t|x_t) \ p(x_t|y_{1:t-1}) dx_{t}}
\label{update1}
\ea
Expressions (\ref{predict}) and (\ref{update1}) are the essential steps
in sequential data assimilation. Using the last update (the posterior,
also often called analysis) as initial condition, the Chapman-Kolmogorov
(prediction) equation (\ref{predict})
is used to forecast the state at the next time
step. When observations $y_t$ become available, they are assimilated
into the model forecast by the update equation (\ref{update1}). 
This cycle constitutes
sequential data assimilation of state-space or Hidden Markov models. The
problem appears in other research fields under different guises,
e.g. Bayesian, optimal, nonlinear or stochastic filtering, or online
inference and learning \citep{Doucet-et-al2001, Cappe-et-al2005}. 


In general, there may be unknown parameters in the model forecast distribution that need to be estimated. We assume that the parameters of the conditional data likelihood are known, since they should be characterized by the measurement process and its associated uncertainties. Several parameter estimation techniques exist; we will focus on maximizing the marginal complete data likelihood, the denominator in Bayes' theorem: 

\be
\text{Marginal complete data likelihood:}    \hskip 0.5cm p(y_{1:t})= {\int p(y_{1:t}|x_{0:t}) \ p(x_{0:t}) dx_{0:t}}
\label{eq:LL} 
\ee
Equation (\ref{eq:LL}) provides a measure of how successfully a particular model is explaining the data. The marginal complete data likelihood is the analog of the traditional likelihood function, but generalized to noisy observational data. This, in turn, implies that different models may be compared and tested for their consistency with observed data, while explicitly acknowledging data uncertainties. In other words, (earthquake) forecasts may be evaluated based on this measure. 


Only in very special cases are the prediction and update equations
(\ref{predict}) and (\ref{update1}) amenable to analytical solutions. In
the case of a linear Gaussian state-space model, the widespread Kalman
filter \citep{Kalman1960, KalmanBucy1961} calculates exactly the
posterior distributions. Much of filtering theory and data assimilation
has been concerned with identifying useful, suitable and computationally
inexpensive filters for a variety of particular problems. For instance,
the extended Kalman filter performs a local tangent linearization of
nonlinear model and observation operators for nonlinear problems. The
Kalman-L\'{e}vy filter \citep{SornetteIde2001}  generalizes the Kalman
filter to L\'{e}vy-law  and power-law distributed model and data
uncertainties. In other cases, numerical integration may be possible, or
approximate grid-based methods, e.g. HMM filters, may be convenient. 
The ensemble Kalman filter \citep{Evensen1994} 
is a Monte Carlo approach to the nonlinear extension of
the Kalman filter by introducing an ensemble of particles with equal
weights, each evolved individually, to approximate distributions. The
general, nonlinear, non-Gaussian, sequential Bayesian estimation
problem, however, seems best solved with sequential Monte Carlo methods
whenever the model's dimensionality is small.  
 
\subsection{Renewal Processes as Forecast Models}
\label{sec:Renew}

Data assimilation is an iterative method that involves two steps, forecast
 (\ref{predict}) and analysis (\ref{update1}), in each cycle.
To formulate the data assimilation problem for earthquakes, we use
a renewal point process as the model in the forecast.
Renewal point processes are characterized by intervals between
successive events that are identically and independently distributed
according to a probability density function that defines the process
\citep{DaleyVereJones2004}. Examples of such a probability density
function (d.f.) include the lognormal d.f., the exponential d.f., the
gamma d.f., the Brownian passage time  d.f. and the Weibull d.f.  The
time of the next event in a renewal process depends solely on the time
of the last event:  
\be
p(t_k|t_{k-1}) = p(t_k - t_{k-1}) = p(\tau)
\label{eq:renew}
\ee
where $\tau$ is the interval between events.
The time of the event $t_{k}$ corresponds to the model state $x_{k}$
in data assimilation. Renewal point processes provide prior information for the analysis,
 which we will discuss next. 


\subsection{Sequential Monte Carlo Methods}
\label{sec:SMC}
Earthquake statistics clearly violate Gaussian approximations in terms of their temporal, spatial and magnitude occurrences, so much so that approximate algorithms based on local Gaussian approximations (e.g. the extended Kalman filter) are unlikely to produce good results. Furthermore, the continuous state space of seismicity rules out methods in which that space is assumed to be discrete (such as grid-based methods). This leaves us with numerical integration techniques and Monte Carlo methods. The former are  numerically accurate but computationally expensive in problems with medium to high dimensionality. 

Sequential Monte Carlo (SMC) methods bridge the gap between these cost-intensive methods and the methods based on Gaussian approximations. They are a set of simulation-based methods that provide a flexible alternative to computing posterior distributions. They are applicable in very general settings, parallelisable and often relatively easy to implement. Early methods were developed in the 70s, but only with the advent of cheap computational power in the mid 90s did they become a widespread tool. Since then, SMC methods have been applied in target tracking, financial analysis, diagnostic measures of fit, missing data problems, communications and audio engineering, population biology, neuroscience, and many more. SMC methods are also known under the names of particle filters, bootstrap filters, condensation, Monte Carlo filters, interacting particle approximations and survival of the fittest. Good introductions can be found in \citep{Arulampalam-et-al2002, Cappe-et-al2005, Cappe-et-al2007, Doucet-et-al2000, Doucet-et-al2001, Kuensch2001, Liu2001, LiuChen1998} and in Chapter 6 of \citet{deFreitas1999}.

Sequential Monte Carlo filters use the techniques of Monte Carlo sampling, of (sequential) importance sampling and of resampling, which we describe briefly below before defining three particular particle filters. 

\subsubsection{Monte Carlo Sampling}
\label{sec:MC}

In Monte Carlo (MC) simulation \citep{Liu2001, RobertCasella2004}, a set of $N$ weighted ``particles" (samples) $x^{(i)}_{0:t}$ are drawn identically and independently from a distribution, say, a posterior $p(x_{0:t}|y_{1:t})$. Then, an empirical estimate of the distribution is given by
\be
\hat{p}_N(x_{0:t}|y_{1:t})= \frac{1}{N} \sum_i^N \delta_{x^{(i)}_{0:t}}(x_{0:t})
\label{MC}
\ee
where $\delta_{x^{(i)}_{0:t}}(x_{0:t})$ denotes the Dirac mass located at $x^i_{0:t}$. The essential idea of Monte Carlo sampling is to convert an integral into a discrete sum. One is often interested in some function of the posterior distribution, say its expectation, covariance, marginal or another distribution. Estimates of such functions $I(f_t)$ can be obtained from
\be
I_N(f_t)=\int f_t(x_{0:t}) \hat{p}_N(x_{0:t}|y_{1:t}) dx_{0:t} = \frac{1}{N} \sum_i^N f_t (x^{(i)}_{0:t})
\label{MCint}
\ee
This estimate is unbiased. If the posterior variance of $f_t(x_{0:t})$ is finite, say $\sigma^2_{f_t}$, then the variance of $I_N(f_t)$ is equal to $\sigma^2_{f_t}/N$. From the law of large numbers,
\be
I_N(f_t) \xrightarrow[N \to \infty]{a.s.} I(f_t) 
\label{mclim}
\ee
where a.s. denotes almost sure convergence. That is, the probability
that the estimate $I_N(f_t)$ converges to the ``true" value $I(f_t)$
equals one in the limit of infinite number of particles. Furthermore, if
the posterior variance $\sigma^2_{f_t} < \infty$, then a central limit
theorem holds: 
\be
\sqrt{N} (I_N(f_t)- I(f_t)) \xrightarrow[N \to \infty]{\Delta} \mathcal{N}(0, \sigma^2_{f_t})
\label{mcCLT}
\ee
where $\xrightarrow[N \to \infty]{\Delta}$ denotes convergence in distribution and $\mathcal{N}(0, \sigma^2_{f_t})$ is the normal (Gaussian) distribution with mean zero and variance $\sigma^2_{f_t}$. The advantage of this perfect Monte Carlo method is therefore that the rate of convergence of the MC estimate is independent of the dimension of the integrand. This stands in contrast to any deterministic numerical integration method, whose rate of convergence decreases with the dimensionality of the integrand. 

Unfortunately, because the posterior distribution is usually highly complex, multi-dimensional and only known up to a normalizing constant, it is often impossible to sample directly from the posterior. One very successful solution for generating samples from such distributions is Markov Chain Monte Carlo (MCMC). Its key idea is to generate samples from a proposal distribution, different from the posterior, and then to cause the proposal samples to migrate, so that their final distribution is the target distribution. The migration of the samples is caused by the transition probabilities of a Markov chain (see, e.g.,  Appendix D of \citet{deFreitas1999}). However, MCMC are iterative algorithms unsuited to sequential estimation problems and will not be pursued here. Rather, SMC methods primarily rely on a sequential version of importance sampling. 

\subsubsection{Importance Sampling (IS)}
\label{sec:IS}
Importance Sampling (IS) introduced the idea of generating samples from
a known, easy-to-sample probability density function (pdf) $q(x)$,
called the importance density or proposal density, and then
``correcting" the weights of each sample so that the weighted samples
approximate the desired density. As long as the support of the proposal
density includes the support of the target density, one can make use of
the substitution  
\be
p(x_{0:t}|y_{1:t})= \frac{p(x_{0:t}|y_{1:t})}{q(x_{0:t}|y_{1:t})} q(x_{0:t}|y_{1:t})
\label{sub}
\ee
to obtain the identity
\be
I(f_t) = \frac{\int f_t(x_{0:t}) w (x_{0:t}) q(x_{0:t}|y_{1:t}) dx_{0:t}}{\int w (x_{0:t}) q(x_{0:t}|y_{1:t}) dx_{0:t}} 
\label{mcIS}
\ee
where $w (x_{0:t})$ is known as the importance weight
\be
w (x_{0:t}) =  \frac{p(x_{0:t}|y_{1:t})}{q(x_{0:t}|y_{1:t})}
\label{ISweight}
\ee
Therefore, if one can generate $N$ independently and identically distributed samples $x^{(i)}_{0:t}$ from the importance density $q(x_{0:t}|y_{0:t})$, a Monte Carlo estimate of $I(f_t)$ is given by
\be
\hat{I}_N(f_t)= \frac{\frac{1}{N} \sum_i^N f_t(x^{(i)}_{0:t}) w(x^{(i)}_{0:t}) }{\frac{1}{N} \sum_i^N w(x^{(i)}_{0:t})} = \sum_i^N f_t(x^{(i)}_{0:t}) \tilde{w}^{(i)}_t
\ee
where the normalized importance weights $\tilde{w}^{(i)}_t$ are given by
\be
\tilde{w}^{(i)}_t = \frac{w(x^{(i)}_{0:t})}{\sum_{j=1}^N w(x^{(j)}_{0:t})}
\label{normISweight}
\ee
For finite $N$, the estimate $\hat{I}_N(f_t)$ is biased, as it is the ratio of two estimates. However, it is possible to obtain asymptotic almost sure convergence $\hat{I}_N(f_t) \xrightarrow[N\to \infty]{a.s.} I(f_t)$ and a central limit theorem provided (i) the importance density support contains the posterior density support, and (ii) the expectations of the weights $w_t$ and $w_t f_t^2(x_{0:t})$ exist and are finite \citep{Geweke1989, deFreitas1999}.  

Thus, the posterior density function can be approximated arbitrarily well by the point-mass estimate
\be
\hat{p}(x_{0:t}|y_{1:t}) = \sum_i^N \tilde{w}^{(i)}_t \delta_{x^{(i)}_{0:t}}(x_{0:t})
\label{ISpost}
\ee


\subsubsection{Sequential Importance Sampling (SIS)}
\label{sec:SIS}
In its simplest form, IS is not adequate for sequential
estimation. Whenever new data 
$y_t$ become available, one needs to recompute the importance weights over the entire state sequence. Sequential Importance Sampling (SIS) modifies IS so that it becomes possible to compute an estimate of the posterior without modifying the past simulated trajectories. It requires that the importance density $q(x_{0:t}|y_{1:t})$ at time $t$ admits as marginal distribution at time $t-1$ the importance function $q(x_{0:t-1}|y_{1:t-1})$: 
\be
q(x_{0:t}|y_{1:t})=q(x_{0:t-1}|y_{1:t-1}) q(x_t|x_{0:t-1},y_{1:t})
\label{siseq}
\ee
after iterating, one obtains: 
\be
q(x_{0:t}|y_{1:t})=q(x_{0}) \prod_{k=1}^t q(x_k|x_{0:k-1},y_{1:k})
\label{sisimp}
\ee
Assuming that the state evolves according to a  Markov process and that the observations are conditionally independent given the states, one can obtain
\be
p(x_{0:t})=p(x_0) \prod_{k=1}^t p(x_{k}|x_{k-1}) \hspace{0.5cm} \text{and} \hspace{0.5cm} p(y_{1:t}|x_{0:t})=\prod_{k=1}^t p(y_{k}|x_{k})
\label{sdf}
\ee
Substituting (\ref{sisimp}) and (\ref{sdf}) into (\ref{normISweight}) and using Bayes' theorem, we arrive at a recursive estimate of the importance weights
\be
\tilde{w}^{(i)}_t \propto \tilde{w}^{(i)}_{t-1} \frac{p(y_t|x^{(i)}_t) p(x^{(i)}_t|x^{(i)}_{t-1} )}{q(x^{(i)}_t|x^{(i)}_{0:t-1}, y_{1:t})}
\label{smcweights}
\ee
where the normalization is provided by $\sum_{j=1}^N
\tilde{w}^{(j)}_t$. Equation (\ref{smcweights}) provides a mechanism for
sequentially updating the importance weights. In summary, SIS provides a
method to approximate the posterior density function (\ref{ISpost}) (or
some function thereof) sequentially in time without having to draw
samples directly from the posterior. All that is required is (i)
sampling from the importance density and evaluating it up to some
constant, (ii) evaluating the likelihood $p(y_t|x^{(i)}_t)$ up to some
proportionality constant, (iii) evaluating the forecast
$p(x^{(i)}_t|x^{(i)}_{t-1} )$ up to some constant, and (iv) normalizing
the importance weights via $\sum_{j=1}^N \tilde{w}^{(j)}_t$.  The SIS
thus makes sequential Bayesian estimation feasible.



\subsubsection{Choice of the Importance Density and Resampling}
\label{sec:impdens}

The problem encountered by the SIS method is that, as $t$ increases, the distribution of the importance weights becomes more and more skewed. For instance, if the support of the importance density is broader than the posterior density, then some particles will have their weights set to zero in the update stage. But even if the supports coincide exactly, many particles will over time decrease in weight so that after a few time steps, only a few lucky survivors have significant weights, while a large computational effort is spent on propagating unimportant particles. It has been shown that the variance of the weights can only increase over time, thus it is impossible to overcome the degeneracy problem \citep{Kong-et-al1994}. Two solutions exist to minimize this problem: (i) a good choice of the importance density, and (ii) resampling. 

\begin{itemize}
\item \textbf{Importance Density}: The optimal importance density is given by: 
\be
q_{opt}(x_t|x_{0:t-1},y_{1:t}) = p(x_t|x_{0:t-1},y_{1:t})= \frac{p(y_t|x_t, x^{(i)}_{t-1}) p(x_t|x^{(i)}_{t-1}) }{p(y_t|x^{(i)}_{t-1})}
\label{optdens}
\ee
because it can be proven to minimize the variance of the importance
weights (see  \citet{Kong-et-al1994} and Chapter 6 of
      \citet{deFreitas1999}). 
However, using the optimal importance
density requires the ability to sample from $p(x_t|x^{(i)}_{t-1},y_t)$
and to evaluate the integral over the
new state $p(y_t|x^{(i)}_{t-1})$  \citep{Arulampalam-et-al2002,
Doucet-et-al2001, deFreitas1999}. In many situations, this is
impossible or very difficult, prompting the use of other
importance densities. Perhaps the simplest and most common choice
for the importance density is given by the prior:  
\be
q(x_t|x_{0:t-1},y_{1:t}) = p(x_t|x_{t-1})
\label{priordens}
\ee
which, although resulting in a higher variance of the Monte Carlo estimator, is usually easy to implement. Many other choices are possible, including additional MCMC steps to sample from the importance density and bridging densities and progressive corrections to herd the particles to the important part of the state space \citep{Arulampalam-et-al2002, Doucet-et-al2001, Liu2001}. 

\item \textbf{Resampling}: Even the optimal importance density will lead to this ``degeneracy" of the particles (few important ones and many unimportant ones). One therefore introduces an additional selection or resampling step, in which particles with little weight are eliminated and new particles are sampled in the important regions of the posterior. \citet{deFreitas1999} and \citet{Arulampalam-et-al2002} provide an overview of different resampling methods. 

Resampling introduces its own problems. Since particles are sampled from discrete approximations to density functions, the particles with high weights are statistically selected many times. This leads to a loss of diversity among the particles as the resultant sample will contain many repeated points. This is known as ``sample impoverishment" \citep{Arulampalam-et-al2002} and is severe when the model forecast is very narrow or deterministic. There are various methods to deal with this problem, including sophisticated methods that recalculate past states and weights via a recursion and MCMC methods, the Resample-Move algorithm and the Regularized Particle Filter (RPF). These filters will not be necessary here because of the broad and highly stochastic model forecast.

Because of the additional problems introduced by resampling, it makes sense to resample only when the variance of the weights has decreased appreciably. A suitable measure of degeneracy of an algorithm is the effective sample size $N_{eff}$ introduced by \citet{LiuChen1998} and defined by
\be
N_{eff}= \frac{N}{1+var (w^{\star i}_t)}
\label{neff}
\ee
where $w^{\star i}_t= p(x^{(i)}_t|y_{1:t})/q(x^{(i)}_t|x^{(i)}_{k-1}, y_k)$ is referred to as the true weight. This may not be available, but an estimate $\hat{N}_{eff}$ can be obtained as the inverse of the so-called Participation Ratio \citep{Mezard1987} (or Herfindahl index \citep{Polakoff81,Lovett88}):  
\be
\hat{N}_{eff}=\frac{1}{\sum_{i=1}^N (w^{(i)}_t)^2}
\label{estneff}
\ee
Thus, resampling can be applied when $\hat{N}_{eff}$ falls below a certain threshold $\hat{N}_{eff}<N_{thres}$.

\end{itemize}

\subsection{Particle Filters and their Numerical Algorithms}

\label{sec:PF}
In this section, we define three particle filters, characterized by particular choices for the importance density and the resampling strategy. The presentation and the pseudo-codes closely follow \citet{Arulampalam-et-al2002}. More on particular particle filters can be found in \citep{Arulampalam-et-al2002, deFreitas1999, Doucet-et-al2000, Doucet-et-al2001, Liu2001, Cappe-et-al2005, Cappe-et-al2007}. The filters are:

\begin{enumerate}

\item The Simple Sequential Importance Sampling (SSIS) particle filter: The simplest particle filter, it uses the prior given by equation (\ref{priordens}) as the (sub-optimal) importance density and does not include a resampling step. 

\item The Optimal Sequential Importance Sampling (OSIS) particle filter: This filter improves on the SSIS by using the optimal importance sampling density (\ref{optdens}), but does not include resampling. 

\item The Optimal Sequential Importance Resampling (OSIR) particle filter: This filter improves on the SIS filters by including a resampling step to counteract the degeneracy of particles. The importance density may either be the prior or the optimal importance density. 

\end{enumerate}

In all particle filters, the prior is obtained by random draw for individual
particles using the forecast model, i.e. the
renewal point process defined by (\ref{eq:renew}).

\subsubsection{Simple Sequential Importance Sampling (SSIS) Filter}
\label{sec:SSIS}

The Simple SIS (SSIS) particle filter is characterized by a lack of resampling and by choosing the prior $p(x_t|x^{(i)}_{t-1})$ as the importance density: 
\be
q(x_t|x_{0:t-1},y_{1:t})=p(x_t|x^{(i)}_{t-1})
\label{impdensprior}
\ee

It can be shown \citep{Arulampalam-et-al2002} that the SSIS can be reduced to the pseudo-code given by Algorithm \ref{SSIS}, where the weights are given by:
\be
w^{(i)}_t \propto w^{(i)}_{t-1} p(y_t|x^{(i)}_t)
\label{SSISw}
\ee
where $p(y_t|x^{(i)}_t)$ is simply the likelihood and the weights are normalized by
\be
\tilde{w}^{(i)}_t=\frac{w^{(i)}_t}{\sum_{j=1}^N w^{(j)}_t}
\label{normSSIS}
\ee

\vspace{0.5cm}
\begin{algorithm}
\caption{ \textit{Simple SIS Particle Filter}}
\label{SSIS}
\begin{algorithmic}
\vspace{.3cm}
\STATE  $[\{x^{(i)}_t,w^{(i)}_t \}^N_{i=1}] = $SSIS$[\{x^{(i)}_{t-1},w^{(i)}_{t-1} \}^N_{i=1},y_t]$
\vspace{.3cm}
\FOR{$i$=1 to $N$}
\STATE Draw $x^i_t \sim p(x_t|x^i_{t-1})$
\STATE Assign the particle a weight, $w^i_t$, according to (\ref{SSISw})
\ENDFOR
\vspace{.3cm}
\end{algorithmic}
\end{algorithm}
\vspace{0.5cm}

This filter is simple and easy to implement. 
However, if the likelihood has a much narrower support than the importance density, then the weights of many particles will be set to zero so that only few active particles are left to approximate the posterior. Depending on the overlap of the support of the two density functions, this particle filter may quickly degenerate in quality. 

\subsubsection{Optimal Sequential Importance Sampling (OSIS) Filter}
\label{sec:OSIS}

The Optimal Simple SIS (OSIS) improves on the SSIS by using the optimal sampling density:
\be
q_{opt}(x_t|x_{0:t-1},y_{1:t}) = p(x_t|x_{0:t-1},y_{1:t})= \frac{p(y_t|x_t, x^{(i)}_{t-1}) p(x_t|x^{(i)}_{t-1}) }{p(y_t|x^{(i)}_{t-1})}
\label{impdensprior2}
\ee
Then, the OSIS filter is implemented by Algorithm \ref{OSISPF}, where the weights are given by substituting the optimal importance density (\ref{impdensprior2}) into the recursive weight equation (\ref{smcweights}) to obtain:
\be
w^{(i)}_t \propto w^{(i)}_{t-1} p(y_t|x^{(i)}_{t-1}) =  w^{(i)}_{t-1} \int p(y_t|x'_t) p(x'_t|x^{(i)}_{k-1}) dx'_t
\label{OSISw}
\ee
Weights are normalized as in equation (\ref{normSSIS}). As was already mentioned, the optimal density suffers from two difficulties: (i) generating samples from the posterior (\ref{impdensprior2}), and (ii) calculating the integral in (\ref{OSISw}). 

\vspace{0.5cm}
\begin{algorithm}
\caption{ \textit{Optimal SIS Particle Filter}}
\label{OSISPF}
\begin{algorithmic}
\vspace{.3cm}
\STATE  $[\{x^{(i)}_t,w^{(i)}_t \}^N_{i=1}] = $OSIS$[\{x^{(i)}_{t-1},w^{(i)}_{t-1} \}^N_{i=1},y_t]$
\vspace{.3cm}
\FOR{$i$=1 to $N$}
\STATE Draw $x^{(i)}_t \sim q_{opt}(x_t|x^{(i)}_{t-1},y_{t}) \propto p(y_t|x_t, x^{(i)}_{t-1}) p(x_t|x^{(i)}_{t-1})$
\STATE Assign the particle a weight, $w^{(i)}_t$, according to (\ref{OSISw})
\ENDFOR
\vspace{.3cm}
\end{algorithmic}
\end{algorithm}
\vspace{0.5cm}

\subsubsection{Optimal Sampling Importance Resampling (OSIR) Filter}
\label{sec:SIR}

To counter the inevitable problem of particle degeneracy, we can use resampling to generate a new set of particles from the (discrete) posterior. Setting the importance density equal to the optimal importance density as in the OSIS particle filter described above, we recover the Optimal Sampling Importance Resampling (OSIR) algorithm given by Algorithm \ref{OSIRPF}. 
 
In the literature, the OSIR filter is usually an implementation with the prior as the (suboptimal) importance density. Such a filter is called the ``bootstrap" filter by \citet{Doucet-et-al2001}. 

\vspace{0.5cm}
\begin{algorithm}
\caption{ \textit{OSIR Particle Filter}}
\label{OSIRPF}
\begin{algorithmic}
\vspace{.3cm}
\STATE  $[\{x^{(i)}_t,w^{(i)}_t \}^N_{i=1}] = $SIR$[\{x^{(i)}_{t-1},w^{(i)}_{t-1} \}^N_{i=1},y_t]$
\vspace{.3cm}
\FOR{$i$=1 to $N$}
\STATE Draw $x^{(i)}_t \sim q_{opt}(x_t|x^{(i)}_{t-1},y_{t}) \propto p(y_t|x_t, x^{(i)}_{t-1}) p(x_t|x^{(i)}_{t-1})$
\STATE Assign the particle a weight, $w^{(i)}_t$, according to (\ref{OSISw})
\ENDFOR
\STATE Calculate total weight: $W=$SUM$[\{w^{(i)}_t\}^N_{i=1}]$
\FOR{$i$=1 to $N$}
\STATE Normalize: $w^{(i)}_t=W^{-1} w^{(i)}_t $
\ENDFOR
\STATE Calculate $\hat{N}_{eff}=\frac{1}{\sum_{i=1}^N (w^{(i)}_t)^2}$
\IF{$\hat{N}_{eff}<N_{thres}$} 
\STATE Resample using Algorithm \ref{algresample}: $[\{x^{(i)}_k,w^{(i)}_t,- \}^N_{i=1}]= $RESAMPLE$[\{x^{(i)}_t,w^{(i)}_t \}^N_{i=1}]$
\ENDIF
\end{algorithmic}
\end{algorithm}
\vspace{0.5cm}

There are many methods to resample from the posterior (see \citet{Doucet-et-al2001} or Chapter 6 of \citet{deFreitas1999} for a discussion of methods, and \citet{Arulampalam-et-al2002} for a brief overview).  The basic idea is to eliminate particles that have small weights and to concentrate on particles with large weights. It involves generating a new set of particles and associated weights by resampling (with replacement) $N$ times from an approximate discrete representation of the posterior. The resulting sample is an independently and identically distributed sample so that the weights are reset to $1/N$. The method of choice of \citet{Arulampalam-et-al2002} is systematic resampling since ``it is easy to implement, takes $O(N)$ time and minimizes the Monte Carlo variation." Its operation is described in Algorithm \ref{algresample}, where $U[a,b]$ is the uniform distribution on the interval $[a,b]$. For each resampled particle $x_t^{j\star}$, this resampling algorithm also stores the index of its parents, which is denoted $i^j$. This is unnecessary and can easily be suppressed, but may be useful in some situations. 

\vspace{0.5cm}
\begin{algorithm}
\caption{ \textit{Systematic Resampling}}
\label{algresample}
\begin{algorithmic}
\vspace{.3cm}
\STATE  $[\{x^{(j \star)}_t,w^{(j)}_t,i^j \}^N_{j=1}] = $RESAMPLE$[\{x^{(i)}_{t},w^{(i)}_{t} \}^N_{i=1}]$
\vspace{.3cm}
\STATE Initialize the CDF: $c_1=0$
\FOR{$i$=2 to $N$}
\STATE Construct CDF: $c_i=c_{i-1}+w^{(i)}_t$
\ENDFOR
\STATE Start at the bottom of the CDF: $i=1$
\STATE Draw a starting point: $u_1\sim U[0,N^{-1}]$
\FOR{$j$=1 to $N$}
\STATE Move along the CDF: $u_j=u_{1}+ N^{-1}(j-1)$
\WHILE{$u_j>c_i$}
\STATE $i=i+1$
\ENDWHILE
\STATE Assign sample: $x^{(j\star)}_t=x^{(i)}_t$
\STATE Assign weight: $w^{(j)}_t=N^{-1}$
\STATE Assign parent: $i^j=i$
\ENDFOR
\end{algorithmic}
\end{algorithm}
\vspace{0.5cm}

While there are of course many more particle filters, each suited to particular applications, we have here presented the standard algorithms. For more advanced particle filters, see for instance \citep{Arulampalam-et-al2002, deFreitas1999, Doucet-et-al2000} and references therein. The particle filters described above will be used below for seismicity models based on point processes. 


\subsection{Parameter Estimation}
\label{sec:PE}

Parameter estimation techniques within sequential Monte Carlo methods are discussed by, e.g., \citep{Doucet-et-al2001, Kuensch2001, Andrieu-et-al2004, Cappe-et-al2005, Cappe-et-al2007}. The methods are either online-sequential or offline-batch methods. For simplicity, we will restrict this section to one particular technique, based on the offline or batch technique of maximizing (an MC estimate of) the complete marginal data likelihood defined in equation (\ref{eq:LL}). The presentation follows \citet{Doucet-et-al2001}.

We assume that both the Markov transition kernel and the conditional data likelihood, defined by equations (\ref{eq:model}) and (\ref{eq:condLL}), respectively, also depend on an unknown, static parameter vector $\theta$. Moreover, we assume the marginal likelihood $L(\theta | y_{1:t} ) = p_{\theta}(y_{1:t} )$ admits a sequential formulation: 

\be
L(\theta | y_{1:t} ) = p_{\theta}(y_{1:t}) = p_{\theta}(y_{0}) \prod_{k=1}^{t} p_{\theta}(y_{k}|y_{0:k-1})
\label{eq:Lseq}
\ee
where the individual predictive likelihoods are defined as
\be
p_{\theta}(y_{k}|y_{0:k-1}) = \int p_{\theta} (y_k, x_k | y_{0:k-1}) dx_k
\label{eq:Lcond}
\ee
These can be estimated from the weighted particles $\{(x_k^{(i,{\theta})}, w_k^{(i,{\theta})}) \}_{1 \leq i \leq N}$ as 
\ba
p_{\theta}(y_{k}|y_{0:k-1}) &=& \int \int p_{\theta}(y_k | x_k) p_{\theta}(x_k | x_{k-1}) p_{\theta}(x_{k-1} | y_{0:k-1}) dx_{k-1} dx_k  \label{eq:Lcondweights} \\
&\approx& \sum_{i=1}^N w_{k-1}^{(i,{\theta})} \int p_{\theta}(y_k | x_k) p_{\theta}(x_k | x_{k-1}^{(i,{\theta})}) dx_k \label{eq:Lapprox1} \\
&\approx& \sum_{i=1}^N w_{k}^{(i,{\theta})}  \label{eq:Lapprox2} 
\ea
where $w_{k}^{(i,{\theta})}$ are the unnormalized weights at the $k^{th}$ time step. 
Expression (\ref{smcweights}) is used to go from the second to the third approximate equality.

The log-likelihood $\ell(\theta)$ is therefore given by
\ba
\ell(\theta)& = &  \log \left( L(\theta | y_{1:t}) \right)  =   \log \left[  \prod_{k=1}^t p_{\theta}(y_{k}|y_{0:k-1})   \right] =  \sum_{k=1}^t  \log \left[  p_{\theta}(y_{k}|y_{0:k-1})   \right]  \nonumber \\
& \approx & \sum_{k=1}^t \log \left[ \sum_{i=1}^N w_{k}^{(i,{\theta})}  \right]
\label{eq:ell}
\ea
Maximizing the sum of the normalized weights given by expression (\ref{eq:ell}) with respect to the parameter set $\theta$ results in the maximum likelihood estimator $\hat{\theta}$:
\be
\hat{\theta} = \rm{arg \ } \rm{max} \left[ \sum_{k=1}^t \log \left( \sum_{i=1}^N w_{k}^{(i,{\theta})}  \right) \right]
\label{eq:MLE}
\ee
 \citet{Doucet-et-al2001, Andrieu-et-al2004, Cappe-et-al2005,
 Cappe-et-al2007} and \citet{OlssonRyden2008} consider the estimator's
 statistical properties. To find the maximum of the log-likelihood in
 equation (\ref{eq:MLE}), one may use the standard optimization
 algorithms, such as gradient-based approaches, the
 expectation-maximization algorithm, or random search algorithms such as
 simulated annealing, genetic algorithms, etc (see, e.g.,
 \citet{SambridgeMosegaard2002}). 
In our parameter estimation experiments (see section
\ref{sec:NumExpPE}),
we chose a combination of a coarse direct grid-search method and a pattern
search method to refine the coarse estimate \citep{HookeJeeves1961,
Torczon1997, LewisTorczon1999}.  

When the number of particles is small, the log-likelihood function will suffer from Monte Carlo noise. Techniques exist to smooth the function in order to find the maximum in these situations (see references above). We were able to perform large particle ensemble computations ($10,000$ particles) in our numerical experiments below, so that the Monte Carlo noise was significantly reduced and we did not need to smooth the likelihood function.

\section{Numerical Experiments and Results}
\label{sec:NumExp}

In this section, we present a simple, pedagogical example of earthquake forecasting based on data assimilation. Our model is the 1-dimensional, temporal lognormal renewal process (section \ref{subsec:model}): the simplest point process, which nevertheless draws much interest in earthquake seismology and seismic hazard, as mentioned above. We assume the process is observed in noise, i.e. the unobservable true occurrence times are perturbed by (additive) identically and independently distributed noise (section \ref{subsec:obs}). The aim of this section is to show an example of how data assimilation provides better forecasts, as measured by the likelihood gain, than a forecast  (``the benchmark") which ignores the data errors (assumes the observed times are the true times). We will compare the performance of the three particle filters defined in section \ref{sec:PF} against each other, and measure their skill against the benchmark (section \ref{sec:NumExpFiltering}). Finally, in section \ref{sec:NumExpPE} we will use maximum likelihood estimation to obtain parameter estimates using both the particle filters and the benchmark. The results in this section thereby demonstrate that data assimilation can help make earthquake forecasting and validating robust with respect to observational data errors.

\subsection{Experiment Design}
\label{sec:NumExpSetUp}

\subsubsection{The Forecast Model: Lognormal Renewal Process}
\label{subsec:model}

Motivated by its relevance to paleoseismology, seismic hazard and the
characteristic earthquake debate, we use a lognormal renewal process as
our forecast model of intervals $\tau$ between subsequent earthquakes 
(section \ref{sec:Renew}):
\be
f_{lognormal}(\tau;\mu, \sigma)  =  \frac{1}{\tau \sqrt{2 \pi} \sigma } \exp(-(\log \tau - \mu  )^2/2\sigma^2) 
\label{lnrp}
\ee
where the parameters $\mu$ and $\sigma$ may need to be estimated. 
In the notation of section \ref{sec:method}, using a physically
meaningful $t_{k}$ for the state variable instead of $x_{k}$, 
the lognormal distribution of the intervals is the transition kernel of
the HMM defined in equation (\ref{eq:model}): 
\be
p\left(t_k|t_{k-1};\mu, \sigma \right)  =  \frac{1}{(t_k-t_{k-1}) \sqrt{2 \pi} \sigma } \exp\left(-\left(\log  \left(t_k-t_{k-1}\right) - \mu  \right)^2/2\sigma^2\right) 
\label{eq:kernel}
\ee
For our pedagogical example, we set the parameters to
\ba
\mu = 1 {\hskip 0.5cm} \text{and} {\hskip 0.5cm} \sigma = \frac{1}{8} 
\label{eq:parmsLn}
\ea
Figure \ref{fig:parms} shows the lognormal distribution 
(solid black curve)
with these parameter values. 

\subsubsection{The Observations: Noisy Occurrence Times}
\label{subsec:obs}

\begin{figure}
\centering
\includegraphics[draft=\IsDraft,width=\halfwidth,keepaspectratio=true,clip]{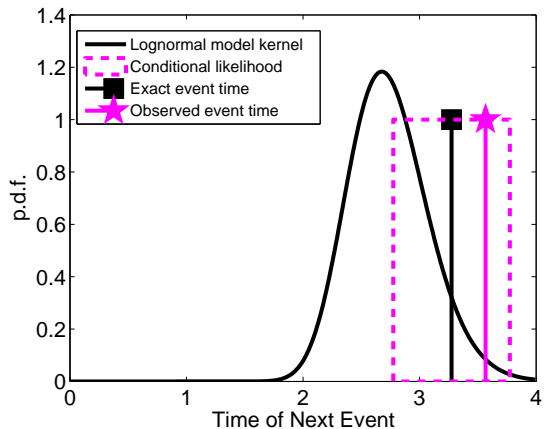}
\caption{Visual comparison of (i) the model transition kernel of the (unobservable) true occurrence times, i.e. the lognormal distribution with parameters $\mu=1$ and $\sigma = 1/8$ (solid black curve), and (ii) the conditional likelihood of the noisy observed occurrence time given the true occurrence time, i.e. the uniform density of width $\Delta=1$ (dashed rectangular magenta line). Also shown are a sample true occurrence time (black square) and a sample observation (magenta star).}
\label{fig:parms}
\end{figure}

We suppose that the $k$-th observed occurrence time $t^o_k$ is a noisy perturbation of the ``true" occurrence time $t^t_k$: 
\be
t^o_k=t^t_k+\epsilon_k
\label{times}
\ee
where $\epsilon$ is an additive noise term distributed according to some distribution $p_{\epsilon}(\epsilon)$. For our numerical experiments below, we choose for simplicity the uniform distribution:
\ba
p_{\epsilon}(\epsilon)= \frac{1}{\Delta} H(\epsilon + \frac{\Delta}{2}) H(\ \frac{\Delta}{2}-\epsilon) 
=  \left\{ \begin{array} {r@{\quad \quad}l}
 \frac{1}{\Delta} &- \frac{\Delta}{2} \leq \epsilon \leq + \frac{\Delta}{2} 
\\  0  & \rm{otherwise} \end{array} \right.
\label{pdfe}
\ea
where $H(\cdot)$ is the Heaviside step function. Substituting $\epsilon=t^o-t^t$ gives the density (conditional likelihood) of the data given the true occurrence time, defined by equation (\ref{eq:condLL}):
\ba 
p^o_{\epsilon}(\epsilon_k) = p(t^o_k|t^t_k) = p(t^o_k - t^t_k)  
=  \left\{ \begin{array} {r@{\quad \quad}l}
 \frac{1}{\Delta} & t^t_k - \frac{\Delta}{2} \leq t^o_k \leq t^t_k + \frac{\Delta}{2} 
\\  0  & \rm{otherwise} \end{array} \right.
\label{lik}
\ea
We set the parameter to
\be
\Delta = 1
\label{eq:delta}
\ee
Figure \ref{fig:parms} shows the lognormal distribution (the model;
solid black curve) 
and the error distribution 
(the conditional likelihood; dashed rectangular magenta line)
for the parameters defined by expressions (\ref{eq:parmsLn}) and (\ref{eq:delta}).

\subsubsection{Initial Condition and Observation Period}
\label{subsec:IC}

We assume for simplicity that the period $T=[a,b]$ over which the point process is observed begins with an event at $t_0=0=a$. We further assume that the true and observed occurrence times of this first event coincide, so that our initial condition $p(x_0)$ is a delta function at $p(x_0) = \delta (t_o=0)$. This assumption can be relaxed: \citet{Ogata1999} provided the relevant equations. One can even predict backwards in time to estimate the time of the last event before the observation period began. 

We assume that the observation period ends with the last observed event $t^o_n=b$. Again, an open interval after the last observed event can be added if needed \citep{Ogata1999}.

\subsubsection{Simulation Procedure}

In this entirely simulated example, we begin by generating the ``true" (unobservable) process. We generate $n$ random samples from the lognormal distribution given by equation (\ref{eq:kernel}) to obtain the sequence of true event times $\{t^t_k \}_{0\leq k \leq n}$. Next, we simulate the observed process by generating $n$ random samples from the uniform conditional likelihood given by equation (\ref{lik}) to obtain the sequence of observed event times $\{t^o_k \}_{0\leq k \leq n}$. 

To perform the particle filtering, we initialize $N$ particles at the
exactly known $t_0=0$. To forecast $t_1$, we propagate each particle
through the model kernel (\ref{eq:kernel}). Given the observation
$t^o_1$ and the model forecast, we use one of the particle filters
described in section \ref{sec:PF} to obtain the analysis/posterior of
$t_1$. The weighted particle approximation of the posterior is then used
to forecast $t_2$ according to equation (\ref{predict}). This cycle is
repeated until the posteriors of all $n$ occurrence times are computed.  

\subsubsection{Optimal Importance Density}

Our simple choice of the uniform conditional likelihood given by
equation (\ref{lik}) allows us to use the optimal importance density
defined in equation (\ref{impdensprior2}) to sample from the
posterior. As stated in section \ref{sec:OSIS}, we need to overcome two
difficulties in order to use the optimal importance density: (i) to
sample from expression (\ref{impdensprior2}), and (ii) to calculate the
integral in (\ref{OSISw}). To sample from (\ref{impdensprior2}), we generate a uniformly distributed random
variable in the bounded interval defined by the quantiles that
correspond to $t^t_k=t^o_k - \Delta/2$ and $t^t_k=t^o_k +
\Delta/2$. 
Then, we invert the prior cumulative distribution function 
$P(t_k |t_{k-1}^{(i)}) \equiv {\rm Pr}({\rm next \ occurrence \ time} \leq t_k|t_{k-1}^{(i)})$ to find a sample $t_k^{(i)}$.
In other words, for each particle $t_{k-1}^{(i)}$, we generate samples from the lognormal distribution $p(t_k|t_{k-1}^{(i)})$, but only in the interval $t_k \in {[t^o_k - \Delta/2, t^o_k + \Delta/2]}$ as permitted by the boxcar likelihood function. Secondly, the integral $p(y_k|t^{(i)}_{k-1})$ can be transformed into error functions, since we need to integrate the lognormal distribution over a bounded interval defined by the boxcar likelihood. The error functions, in turn, can be easily calculated by standard numerical functions.

\subsection{Particle Filtering}
\label{sec:NumExpFiltering}

This section presents examples of the forecast and posterior
distributions 
using a large number of particles ($N=10,000$).
We also compare the performance of the three particle
filters SSIS, OSIS, and OSIR, defined in section \ref{sec:PF},
against each other and against the benchmark, which entirely
neglects the presence of data uncertainties. We assume in this section
that the parameters are known. 

\subsubsection{Forecasts}
\label{subsec: Forecast}
\begin{figure}
\centering
\includegraphics[draft=\IsDraft,width=\halfwidth,keepaspectratio=true,clip]{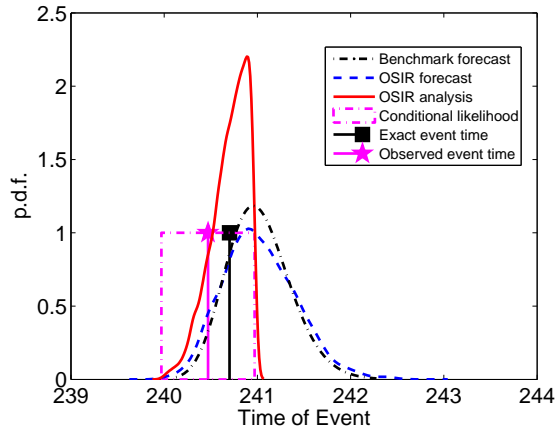}
\caption{Illustration of the OSIR particle filter for one particular event in the simulated process: (i) The OSIR forecast (blue dashed) takes into account both the lognormal transition kernel and the uncertainty in the last occurrence time, thus resulting in a broader forecast than the benchmark's forecast (black dash-dotted), which ignores this uncertainty; (ii) the conditional likelihood (magenta dash-dotted rectangular), and (iii) the resulting posterior distribution (red solid). Also shown are the (unobservable) ``true" event (black square) and the observed event (magenta star). The distributions shown here are kernel density estimates derived from the weighted particles of the particle filter. 
}
\label{fig:Forecast}
\end{figure}

Figure \ref{fig:Forecast} shows an example of the forecast distributions
of the benchmark and of the OSIR particle filter. The benchmark assumes that
the observed events correspond to the 
previous ``true" events without data
errors. Therefore, the forecast distribution of the next occurrence time
is simply the lognormal distribution, as opposed to the particle filter
forecasts, which are broadened due to the integration of the lognormal
distribution over the uncertainty in the last occurrence time. As a
result, the benchmark forecast is artificially sharper than the filter
forecasts. In some cases, the sharper peak may lead to higher
likelihoods -- but the benchmark will pay heavily in the case 
when the observed event is in the tails of its sharp forecast. In 
those cases, the broaden particle filter forecasts will
more than recuperate. Section \ref{subsec:Performance} compares the
likelihood scores and gains of the benchmark and the particle filters.

\subsubsection{Posteriors}
\label{subsec: Posteriors}

Figure \ref{fig:Forecast} also shows Bayes' theorem at work using the OSIR filter: The weighted particle approximation of the forecast (dashed blue) is combined with the conditional likelihood (dash-dotted magenta) to generate an estimate of the posterior of the true occurrence time (solid red).  The posterior (or analysis) is the best possible estimate of the true occurrence time, given a priori model information and the data. The posterior is used as the initial condition for the next forecast. 

Note that the posterior's support is limited by the boxcar data likelihood function. However, because of the finite width of the kernel we chose to represent the weighted particles, the posterior seems to extend just beyond the sharp edges. More sophisticated kernels should be able to solve this minor visualization issue.

\subsubsection{Comparison of Particle Filters}
\label{subsec: Comparison}

\begin{figure}
\centering
\includegraphics[draft=\IsDraft,width=\halfwidth,keepaspectratio=true,clip]{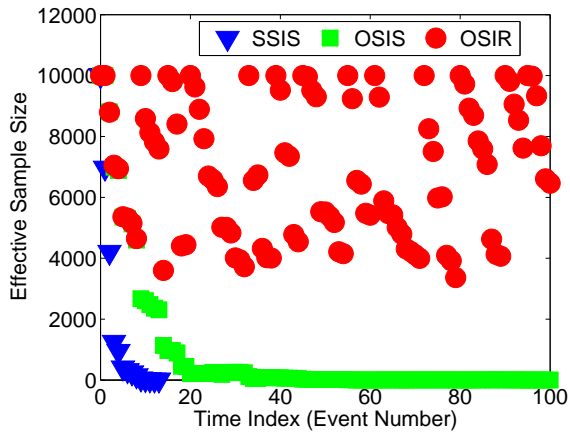}
\caption{The evolution of the 
estimated effective sample size $\hat{N}_{eff}$, a measure of the skewness of the particle weight distribution, as a function of time in a simulation of 100 events: the SSIS (blue triangles) quickly collapses; the OSIS (green squares) survives indefinitely but has a strongly skewed weight distribution; the OSIR (red circles) is able to rejuvenate its particles by resampling.   
 }
\label{fig:ESS}
\end{figure}

As discussed in section \ref{sec:impdens}, particle filters without
resampling will eventually collapse, because of a highly skewed
distribution of weights: a few particles will carry most of the weight
while the majority carry little or no weight. Figure \ref{fig:ESS} shows
the evolution of the 
estimated effective sample size  ($\hat{N}_{eff}$),
 the measure of the skewness
of the weight distribution defined in equation (\ref{estneff}), for a
simulation of 100 events, for each of the three filters.   

\begin{figure}
\centering
\includegraphics[draft=\IsDraft,width=\halfwidth,keepaspectratio=true,clip]{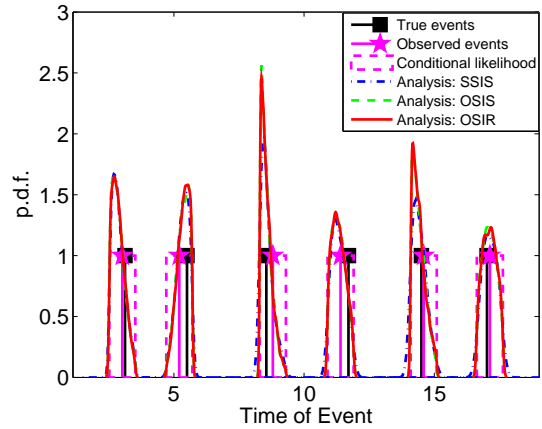}
\caption{Kernel density visualization of the weighted particle approximations of the posterior distributions of the unobservable ``true" occurrence times: comparison of the three particle filters (SSIS in dash-dotted  blue, OSIS in dashed green, OSIR in solid red) for events 1 to 5. Also shown are the conditional data likelihoods (rectangular dashed magenta lines) and the ``true" times (black squares) and the observed times (magenta stars). 
 }
\label{fig:Steps15}
\end{figure}

\begin{figure}
\centering
\includegraphics[draft=\IsDraft,width=\halfwidth,keepaspectratio=true,clip]{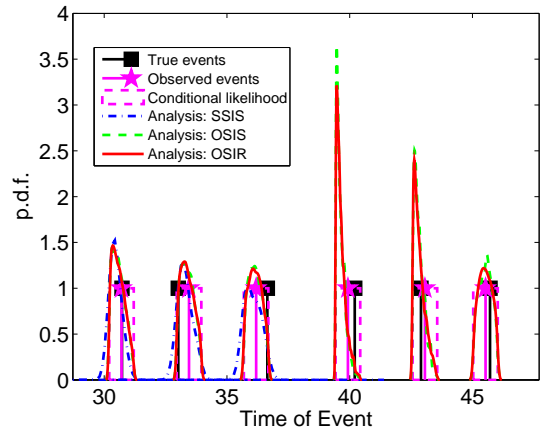}
\caption{Same as Figure \ref{fig:Steps15}, but for events 16 to 20. The death of the SSIS filter occurs after event 17. }
\label{fig:Steps1620}
\end{figure}

Since the SSIS uses the model prior as sampling density, 
many of the particles fall outside the boxcar
likelihood and their weights are set to zero, according to equation (\ref{SSISw}).
Thus, fewer and fewer particles are used to represent the posterior,
resulting in a deteriorating representation of the posterior
distributions. Figure \ref{fig:Steps15} shows kernel density
visualizations of the weighted particle approximations of the posterior
distributions of the first 5 unobservable ``true" occurrence times. The
SSIS shows signs of departures from the OSIS and the OSIR quickly, until
the SSIS filter finally ``dies" after  event 17 
(see Figure \ref{fig:Steps1620}). 
Nevertheless, the SSIS offers the advantage of a very simple filter, which can be used effectively for a few time steps. 

\begin{figure}
\centering
\includegraphics[draft=\IsDraft,width=\halfwidth,keepaspectratio=true,clip]{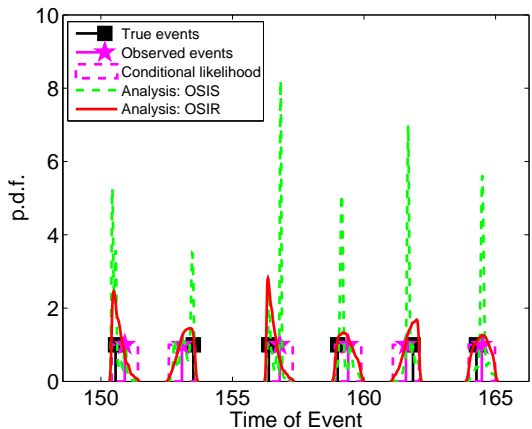}
\caption{Same as Figure \ref{fig:Steps15}, but for events 500 to 505. The OSIS filter shows progressively worse performance. }
\label{fig:Steps500505}
\end{figure}

In the OSIS filter, particles are propagated using the optimal
importance density,  e.g., equation (\ref{OSISw}),
so that their weights cannot be set to zero as in the SSIS filter.
 Nevertheless, the increasingly skewed distribution of the
particle weights leads to poor representations of the posteriors. Figure
\ref{fig:Steps1620} shows first signs of problems after event 16: spikes in the distribution belie the presence of ``heavy-weight"
particles. Figure \ref{fig:Steps500505} shows the posteriors of the true
occurrence times for events 500 to 505. A few heavy-weights completely
dominate the weight distribution, resulting in spikes in the
posterior. Since the particles never ``die", the OSIS filter survives
indefinitely, but the posterior representations progressively worsen as
measured by the effective sample size in Figure
\ref{fig:ESS}. Nevertheless, the OSIS filter, through its use of the
optimal importance density, is a significant improvement on the SSIS
filter and may be very useful for small data sets.  

\begin{figure}
\centering
\includegraphics[draft=\IsDraft,width=\halfwidth,keepaspectratio=true,clip]{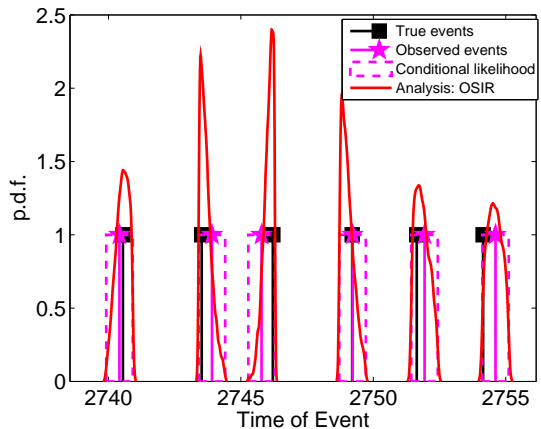}
\caption{Same as Figure \ref{fig:Steps15}, but for events 1000 to 1005. The OSIR filter continues to perform well.   }
\label{fig:Steps10001005}
\end{figure}

The OSIR filter re-samples the particles to equalize the particle
weights whenever the  estimated
effective sample size falls below a certain threshold,
i.e., $\hat{N}_{eff}<N_{thres}$, as described in section \ref{sec:SIR}. 
In our experiments, we chose 
$N_{thres}$  to be one third of the particle number $N$. 
The effective sample size, and thus the filter, can therefore be rejuvenated
(see Figure \ref{fig:ESS}) to continue to provide good approximations of
the posteriors (see Figures \ref{fig:ESS} to
\ref{fig:Steps10001005}). Figure \ref{fig:Steps10001005} shows the
posteriors for events 1000 to 1005.  

Theoretically, the posteriors of the three filters should coincide as the number of particles approaches infinity. We have already mentioned the different weight distributions as one reason for the visible differences for a finite number of particles. Another source of differences lies in the inherent Monte Carlo noise, i.e. fluctuations due to the pseudo-random numbers used to generate samples.

\subsubsection{Comparison against the Benchmark: Likelihood Scores and Likelihood Gain}
\label{subsec:Performance}

To measure the improvement of our ``earthquake" forecasts based on data
assimilation over the naive benchmark, which ignores data uncertainties,
we use the log-likelihood score and other common, likelihood-based
measures of earthquake forecasts based on point processes
\citep{DaleyVereJones2004, HarteVereJones2005,  Helmstetter-et-al2006,
Kagan2007c}. However, we extend the measures by taking into account
uncertainties (see also \citep{Ogata1999}), as suggested by
\citet{Doucet-et-al2001, Andrieu-et-al2004, Cappe-et-al2005} for HMM
estimation problems. In particular, we employ the marginal
log-likelihood of the data, defined by equation (\ref{eq:LL}), which
reflects both the model forecast and the conditional likelihood function
(the measurement process). This marginal likelihood is nothing but the
denominator in Bayes' theorem (\ref{bayesDA}), which judges how well the
data are explained, assuming both a model forecast and a measurement
process.  

For the particle filters, the marginal log-likelihood of the data can be approximated by equation (\ref{eq:ell}). The benchmark effectively assumes that the measurement process is perfect, such that the box-car conditional likelihood is replaced by a Dirac function $p(t_k^o|t_k^t)=\delta(t_k^o-t_k^t)$. The benchmark log-likelihood score is thus simply obtained by using the lognormal density function and plugging in the observed occurrence times. Since this is a stochastic prediction problem, it is also of interest to compare these forecasts to the ideal case of having access to the ``true" occurrence times. For the ``true" process, the log-likelihood score is obtained by using the lognormal distribution and plugging in the ``true" event times, again replacing the conditional likelihoods by Dirac functions. Since this will only give the score for one particular realization, we also calculate the average log-likelihood score per event, given by the negative entropy of the lognormal distribution, which is available analytically. In this section, we assume the parameters are known exactly. 

\begin{figure}
\centering
\includegraphics[draft=\IsDraft,width=\halfwidth,keepaspectratio=true,clip]{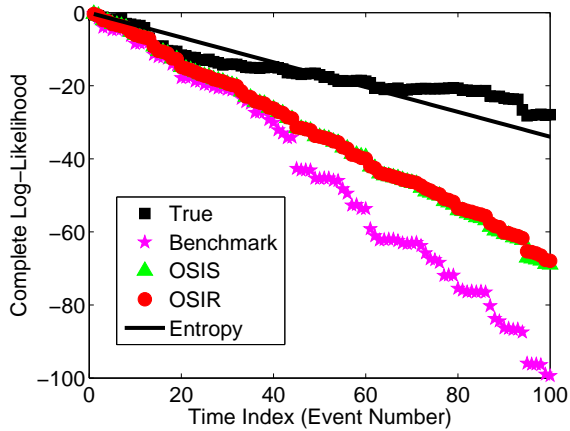}
\caption{ Evolution of the cumulative complete marginal log-likelihood using (i) the OSIS and OSIR particle filters, and (ii) the benchmark. Also shown are the log-likelihood of the ``true" (unobservable) times using the ``true" event times (black), and the average log-likelihood given by the negative entropy of the lognormal distribution, shown as the straight black line. 
 }
\label{fig:LL}
\end{figure}

Figure \ref{fig:LL} shows the evolution of the cumulative (complete) marginal
log-likelihood of the data using the OSIS and OSIR particle filters and the benchmark for a simulation of
100 events (since the SSIS filter collapses quickly, we discard it henceforth). The benchmark clearly performs much worse than the particle filters, i.e. the observed data can be explained much better with the particle filters than with the benchmark. This simple observation demonstrates a major potential benefit of data assimilation techniques for earthquake forecasting: Taking realistic uncertainties into account produces better forecasts and explains the data better than ignoring observational errors.  

All the particle filters should produce the same log-likelihood scores, at least theoretically as the number of particles approaches infinity. However, as discussed above, the SSIS filter quickly collapses, while the OSIS filter suffers from a strongly skewed distribution of particle weights. For this reason, and because of random Monte Carlo noise, there exists a small difference between the OSIS and the OSIR scores. 

To understand how much better the particle filters perform than the benchmark, we also include the log-likelihood scores of the ``true" event times, assuming access to the past ``true" times. The process being stochastic, each realization will give a different score. The average log-likelihood score of the ``true" process, on the other hand, is given by the negative entropy of the lognormal distribution using the ``true" occurrence times. No forecast method could perform better (on average) than the ``true" process. For the particular realization in Figure \ref{fig:LL}, the particle filters obtain scores almost half as negative as the benchmark. 

\begin{figure}
\centering
\includegraphics[draft=\IsDraft,width=\halfwidth,keepaspectratio=true,clip]{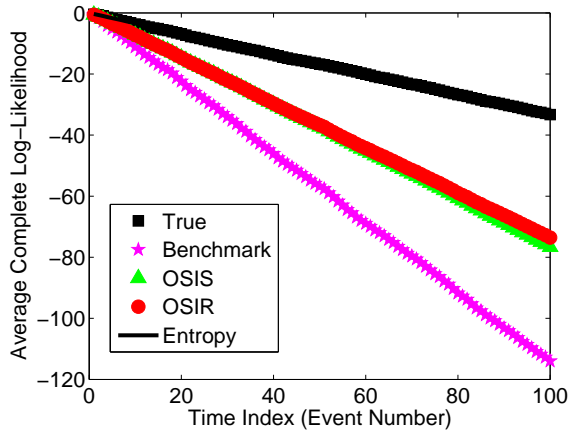}
\caption{ Evolution of the sample mean of the cumulative complete marginal
 log-likelihood, averaged over 100 realizations of a 
100-event point process: 
mean log-likelihood of the ``true" process (black squares), of the particle filters (OSIS - green triangles, OSIR - red circles) and of the benchmark (magenta stars). The negative entropy of the lognormal distribution is shown as the straight black line (obscured by the black squares).
 }
\label{fig:LLavg}
\end{figure}

To investigate the average performance improvement, we simulated 100
realizations of a  100-event point process. We calculated the mean of the
log-likelihood scores at each event index, as shown in Figure
\ref{fig:LLavg}. Fluctuations are now mostly smoothed out. The mean
``true" likelihood scores now match exactly the negative entropy
predictions. The OSIR performs slightly better than the OSIS, for the
reasons stated above, and significantly better than the benchmark. Since
parameters are fixed, the likelihood ratio test would strongly favor the
particle filters.

To measure the quality of a point process forecast with respect to a reference forecast, we employ several common measures. The individual probability gain $G^{(1)}_k$ measures the ratio of the likelihood $p_1(t_k^o)$ of the $k^{th}$ observed event under a specific forecast over the likelihood of the same event under a reference forecast $p_0(t_k^o)$:
\be
G^{(1)}_k = \frac{p_1(t_k^o)}{p_0(t_k^o)}
\label{eq:G}
\ee
The individual probability gain $G^{(1)}_k$ measures how much better the event is explained by our particle filter forecast over the naive benchmark forecast. $G^{(1)}_k=1$ corresponds to no improvement. Since usually log-likelihood scores are used rather than likelihood values, it is common to use the (individual) log-likelihood ratio or log-likelihood gain, defined by:
\be
R^{(1)}_k= \log G^{(1)}_k = \log  \left(\frac{p_1(t_k^o)}{p_0(t_k^o)} \right) = \log p_1(t_k^o) - \log p_0(t_k^o) = LL_1(t_k^o) - LL_0(t_k^o)
\label{eq:R1}
\ee
where $LL(t_k^o)$ is the marginal log-likelihood of event $t_k^o$. Now $R^{(1)}_k=0$ corresponds to no improvement. 

The (cumulative) probability gain $G^{(n)}$ per earthquake of the proposed forecast with respect to a reference forecast is defined as \citep{KaganKnopoff1977, DaleyVereJones2003, DaleyVereJones2004, HarteVereJones2005, Helmstetter-et-al2006, Kagan2007c}:
\be
G^{(n)}= \exp \left( \frac{LL_{1}(n) - LL_{0}(n)}{n}\right) 
\label{eq:Gn}
\ee
where  $LL_1(n)$ and
$LL_0(n)$ are the cumulative marginal log-likelihood scores of the proposed model and a
reference model, respectively, for the $n$ considered events. This measure quantifies the cumulative improvement due to the proposed forecast over a reference forecast. The measure is motivated by its expression as the geometric average of the individual conditionally independent probability gains:
\be
G^{(n)}= \left[ \prod_{k=1}^n G^{(1)}_k \right]^{\frac{1}{n}} =  \left[ \prod_{k=1}^n \frac{p_1(t_k^o)}{p_0(t_k^o)} \right]^{\frac{1}{n}} = \left[ \frac{ \prod_{k=1}^n p_1(t_k^o)}{\prod_{k=1}^n p_0(t_k^o)} \right]^{\frac{1}{n}} 
\label{eq:GnR}
\ee
where the product over all $k=1,...,n$ events specifies the joint probability density of the entire process under a specific model. 

Finally, the (algebraic) average log-likelihood ratio or average log-likelihood gain for $n$ events is defined by
\be
R^{(n)} = \log G^{(n)} = \frac{1}{n} \sum_{k=1}^n R^{(1)}_k =  \frac{1}{n} \sum_{k=1}^n \left( LL_1(k) - LL_0(k) \right)
\label{eq:Rn}
\ee
In our experiments, the benchmark is the reference forecast, i.e. we directly measure any improvement of the particle filters over the benchmark. 

\begin{figure}
\centering
\includegraphics[draft=\IsDraft,width=\halfwidth,keepaspectratio=true,clip]{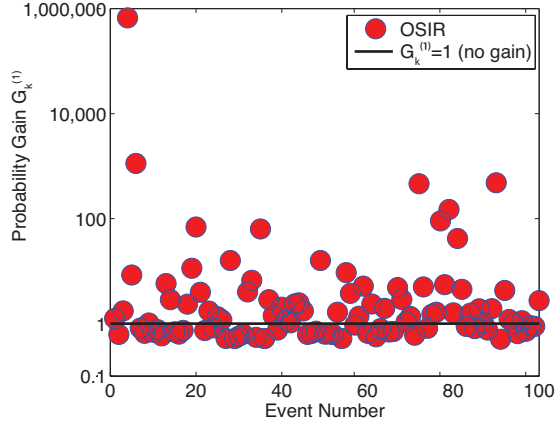}
\caption{Probability gains of the OSIR particle filter over the benchmark for each individual earthquake $k$.
 }
\label{fig:G}
\end{figure}

For a 100-event point-process simulation, we
first calculated the individual probability gains $G^{(1)}_k$ for each event $t^o_k$, as shown in Figure \ref{fig:G}. The individual gains $G^{(1)}_k$ fluctuate wildly, from
about 0.4 to $10^6$. In other words, there are many events that are better forecast by the benchmark than by the OSIR particle filter ($G^{(1)}_k<1$), but there are some events for which the particle filter outperforms the benchmark by a factor of $10^6$. The (cumulative) probability gain $G^{(100)}$ per earthquake is $G^{(100)}=1.60$, i.e. in a geometric average sense, the OSIR particle filter performs better by a factor of 1.6.

The seemingly surprising occurrence of $G^{(1)}_k<1$ forecasts can be explained by the fact that the benchmark forecasts are sharper than the particle filter forecasts, since the benchmark does not take into account the uncertainty in the last occurrence time (compare the forecasts in Figure \ref{fig:Forecast}). As a result, if the next observed event actually falls into the center of the benchmark forecast, the likelihood of the data is higher under the benchmark forecast than under the broadened particle filter forecast. Thus, frequently, the benchmark produces higher likelihood scores than the filters. However, precisely because the benchmark forecast does not take into account data errors, the forecasts are overly optimistic. When the observed events fall outside of this artificially narrow window, the particle filter performs better than the benchmark, and sometimes immensely better. Such surprises for the benchmark are reflected in the very large individual likelihood gains of up to $10^6$. 

\begin{figure}
\centering
\includegraphics[draft=\IsDraft,width=\mediumwidth,keepaspectratio=true,clip]{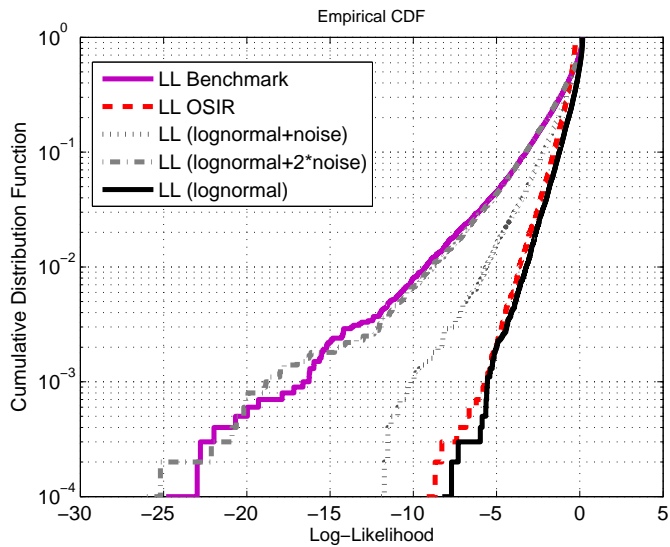}
\caption{Empirical cumulative distribution functions of the log-likelihood scores for each event obtained by the OSIR particle filter, the benchmark, the ``true" process in a 10,000-event point-process simulation. Also shown are two distributions explained in the text. 
 }
\label{fig:LLcdf}
\end{figure}

To illuminate the performance of the OSIR particle filter further, we simulated a 10,000-event point-process and calculated, for each event $k$, the individual log-likelihood scores $LL_1(k)$  and $LL_0(k)$ of the OSIR particle filter and the benchmark, respectively, and their log-likelihood ratios $R^{(1)}_k$. Figure \ref{fig:LLcdf} displays the empirical cumulative distribution functions of the OSIR particle filter's scores (red dashed) and of the benchmark's scores (purple solid). For comparison, we also show the distribution of log-likelihood scores obtained by the using the ``true" process (black solid), and by two other distributions, explained below. The log-likelihood distribution of the ``true" process has consistently the highest scores, up to statistical fluctuations, as expected. The log-likelihood scores of the OSIR particle filter, however, are not consistently better than the benchmark (as already seen in Figure \ref{fig:G}). Rather, the highest scores of the benchmark are higher than those of the particle filter. These values correspond to those events that occur right in the middle of the overly optimistic, overly sharp forecast of the benchmark, thus resulting in a higher score compared with the broadened OSIR particle filter forecast. However, the scores of the benchmark quickly become worse than the particle filter's, and indeed the lowest scores are orders of magnitude lower than the particle filter's. The body and tail of the distributions, emphasized in Figure \ref{fig:LLcdf} by the semi-logarithmic axes, show the particle filter's advantage: the benchmark sometimes produces terrible forecasts, for which it pays with a poor score. At the same time, the individual particle filter's scores always remain relatively close to the scores of the ``true" process. 

We found it helpful to include two other distributions of log-likelihood scores in Figure \ref{fig:LLcdf}. To produce the first distribution, labeled LL(lognormal $+$ noise), we simulated lognormally distributed random variables $x_k$, and then perturbed each by an additive, uniformly distributed error with the same distribution as the observational uncertainty that perturbs the lognormal point process. We then used these random variables and calculated their likelihood using the original lognormal function. For the second distribution, labeled LL(lognormal $+ 2*$noise), we perturbed $x_k$ twice with the same noise and again calculated their log-likelihood under a lognormal function.  The point is to show that the log-likelihood scores of the benchmark naturally come from the fact that we assume a lognormal function in the  
calculation of the likelihood scores, but that the random variables we observe  
are not actually lognormally distributed. In fact, the benchmark makes  
2 mistakes: (i) the origin point (start of the interval) is a  
perturbed version of the last true occurrence time, and (ii) the  
observed next event is again a perturbed version of the next true  
occurrence time. Thus, the interval is perturbed twice. In other  
words, the simulated LL(lognormal +2*noise) should correspond exactly  
to the log-likelihood distribution of the benchmark (up to statistical fluctuations). 

\begin{figure}
\centering
\includegraphics[draft=\IsDraft,width=\mediumwidth,keepaspectratio=true,clip]{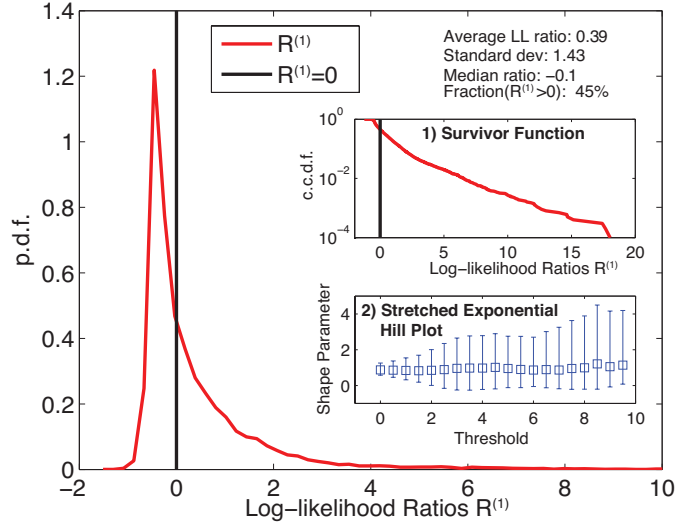}
\caption{ Kernel density estimate of the probability density function of the log-likelihood ratios $R_k^{(1)}$ between log-likelihood scores of the OSIR particle filter and the benchmark. Inset 1: Survivor function. Inset 2: Hill plot of the maximum likelihood estimates of the shape parameter of a stretched exponential distribution along with 95$\%$ confidence bounds, as a function of the threshold above which the parameter is estimated. 
 }
\label{fig:LLSE}
\end{figure}

Figure \ref{fig:LLSE} displays the kernel density estimate of the individual log-likelihood ratios $R_k^{(1)}$, a direct comparison of the OSIR particle filter and the benchmark for each event. The vertical black line at $R^{(1)}=0$ separates the region in which the benchmark performs better ($R_k^{(1)}<0$) from the one in which the particle filter performs better ($R_k^{(1)}>0$). The statistics of this distribution are particularly illuminating: the median is $R^{(1)}=-0.1$, implying that more than $50\%$ of the time, the benchmark outperforms the particle filter (the exact percentage of events for which the benchmark performs better is $55\%$). The most probable value is close to the median value. However, the amount by which the benchmark outperforms the OSIR particle filter is never very much, since the OSIR forecast is never much broader than the benchmark forecast. Thus the potential loss of the OSIR particle filter is limited, as seen by the truncation of the distribution for low log-likelihood ratios. At the same time, the tail of the distribution towards large log-likelihood ratios decays much more slowly. Inset 1 of Figure \ref{fig:LLSE} shows the survivor function in semi-logarithmic axes to emphasize the slow decay. We found that a slightly stretched exponential distribution \citep{LaherrereSornette1998, Sornette2004} fits the tail quite nicely, with a shape parameter (exponent) of about $0.9 \pm0.3$ ($95\%$ confidence bounds) (see Inset 2 of Figure \ref{fig:LLSE} for a stretched exponential Hill plot of estimated shape parameter versus the threshold above which it is estimated). As a result of the stretched exponential tail of the log-likelihood ratio, the potential benefit of the OSIR particle filter can be enormous, while its potential disadvantage is limited. As a result, the average log-likelihood ratio is $R^{(10,000)}=0.39$, despite the negative median. 

One may wonder whether other scores would always favor the OSIR particle filter. For instance, one could declare alarms based on exceeding a certain probability threshold and then evaluate the frequency of hits and misses. The performance could be measured by the Molchan error diagram \citep{Molchan1990, MolchanKagan1992}. It may be possible to design a utility or loss function that actually favors the benchmark. However, whether such a utility function exists and is reasonable depends on the application. Most common measures should favor the OSIR particle filter, since, given the uncertainty in the observed times, the filter's formulation takes into account all available information, while the benchmark ignores the knowledge of a measurement process. 

\subsection{Parameter Estimation}
\label{sec:NumExpPE}

So far, we have assumed that the parameters of the lognormal
distribution are known. In reality, one would like to estimate the
parameters from the observed occurrence times. As stated in section
\ref{sec:PE}, in this article we perform offline maximum likelihood
estimation of a batch of data at a time. Online, sequential parameter
estimation is discussed by, e.g., \citet{Doucet-et-al2001, Kuensch2001,
Andrieu-et-al2004, Cappe-et-al2005, Cappe-et-al2007}. In particular, we
maximize the complete marginal data likelihood, approximated by equation
(\ref{eq:ell}), to estimate parameters. To find the maximum, we first
perform a coarse grid-search over the parameter space and then use a
a pattern search algorithm \citep{HookeJeeves1961, Torczon1997, LewisTorczon1999}. In this section, we
first describe the estimation and compare the parameter estimates of the OSIR particle filter with those of the 
benchmark for single simulations of 
either 200 events or 10 events. 
We then show results for a large number of simulations to demonstrate
statistical bias in the estimates of the benchmark and much better performance of 
the OSIR. 

\begin{figure}
\centering
\includegraphics[draft=\IsDraft,width=\mediumwidth,keepaspectratio=true,clip]{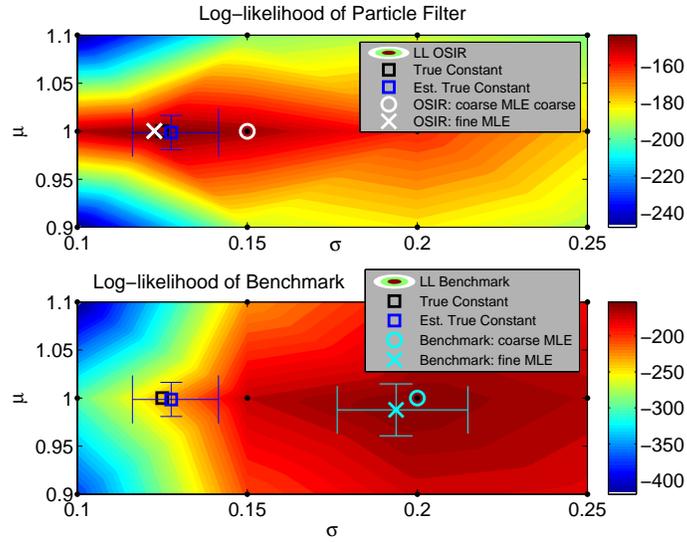}
\caption{Top panel: Marginal data log-likelihood contours of a simulation of a 200-event point process,
approximated by the OSIR likelihood equation (\ref{eq:ell}), as a function of the parameters $\theta=\{\mu, \sigma\}$. We also show (i) the ``true" constants  (black square), (ii) the maximum likelihood estimates (and $95\%$ confidence bounds) using the (unobservable) ``true" occurrence times (blue square), (iii) the coarse search-grid (black dots), (iv) the maximum of this coarse grid (white circle), and (v) the final OSIR maximum likelihood estimate (white cross), attained via the pattern search method. Bottom Panel: As above, but using the benchmark log-likelihood function, along with coarse-grid maximum (cyan circle) and the final maximum likelihood estimate of the benchmark (cyan cross with $95\%$ confidence bounds). 
 }
\label{fig:LLbm}
\end{figure}

To demonstrate the method, we simulated 200 events and then performed
the combined grid-search and pattern search optimization to estimate
parameters. The top panel in Figure \ref{fig:LLbm} shows approximate contours of the
log-likelihood as a function of the two parameters $\mu$ and
$\sigma$. The small black circles show the locations of the coarse
grid-search. The white circle marks the maximum given by the
grid-search. From here, the pattern-search method refined the estimates
iteratively 
until it converged at the white cross,
which thus marks the maximum likelihood estimate using the OSIR particle filter.

To judge the quality of the OSIR 
estimate, we compare its location (white cross) with the ``true" constants used for the simulation (black square) and the maximum likelihood estimates based on the ``true" occurrence times (blue square with 95$\%$ confidence bounds). The estimates nearly overlap and all fall within the confidence bounds of the estimate based on the (unobservable) ``true" occurrence times. 
 
In contrast, the benchmark estimate is clearly biased in the parameter
$\sigma$, as can be seen in the bottom panel of Figure \ref{fig:LLbm}: its 95$\%$ confidence bounds do not overlap with the
target. At the same time, its estimate of $\mu$ seems consistent at this
level of resolution. The benchmark attempts to counter-act its failure to 
account for the occurrence of large errors
in observed occurrence times by increasing the variability (the
variance) of the lognormal process with respect to the actual
value. However, the increased breadth of the benchmark forecast on
average decreases its likelihood score. The benchmark cannot attain the
same likelihood values as the OSIR,
even by attempting to counter-act the occurrence of large errors
in observed occurrence times. 
The likelihood score of the estimates using the exact
occurrence times (blue square) is naturally much higher than the  OSIR 
estimate (white cross) which does not have access to the exact occurrence times. The OSIR's likelihood in turn is much higher than the benchmark
estimate (cyan cross). The likelihood function of the benchmark is
different from the OSIR  likelihood function, of course, as seen in the bottom panel of Figure \ref{fig:LLbm}. 

\begin{figure}
\centering
\includegraphics[draft=\IsDraft,width=\mediumwidth,keepaspectratio=true,clip]{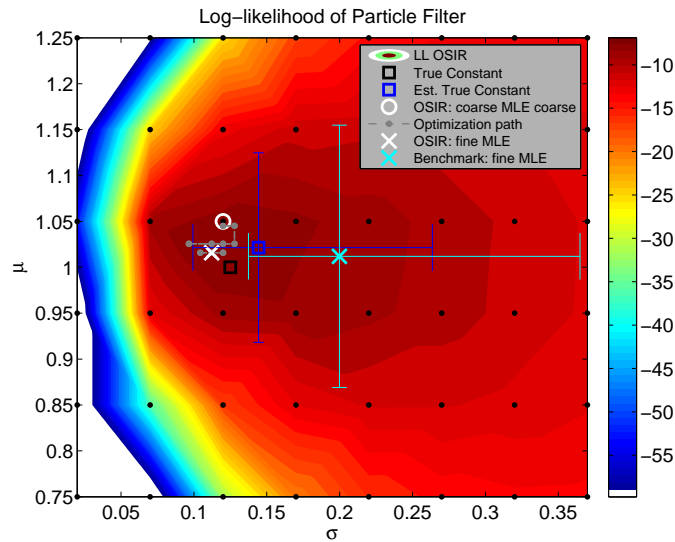}
\caption{Same as top panel in Figure \ref{fig:LLbm} but for a simulation of only 10 
events of the point process.
 }
\label{fig:LL10}
\end{figure}

We have also estimated parameters in simulations of few events. Figure
\ref{fig:LL10} shows the approximate log-likelihood contours of the
OSIR particle filter for a simulation of only 
10 events of the point process. Such a small sample size mimics good
paleoseismic data sets. The benchmark
estimate is now much closer to the ``true" constant (black square), but
the benchmark's 95$\%$ confidence interval still does not include the
target. However, the maximum likelihood estimate (blue square) using the
exact event times is within that interval, and its own confidence
interval also includes the benchmark estimate. Nevertheless, the
OSIR maximum likelihood estimate remains a much better estimator than the benchmark. 

To establish that the OSIR particle filter
consistently obtains better
parameter estimates than the benchmark, we investigated the
statistics of repeated estimates from different realizations of the 
point process. Desirable properties of parameter estimators include
unbiasedness, consistency and (asymptotic) normality. We refer to the
references in the beginning of this section and to
\citet{OlssonRyden2008} and references therein for theoretical work on
the properties of the
maximum likelihood estimator of the  OSIR particle filter. We concentrate on demonstrations by simulation. 

\begin{figure}
\centering
\includegraphics[draft=\IsDraft,width=\mediumwidth,keepaspectratio=true,clip]{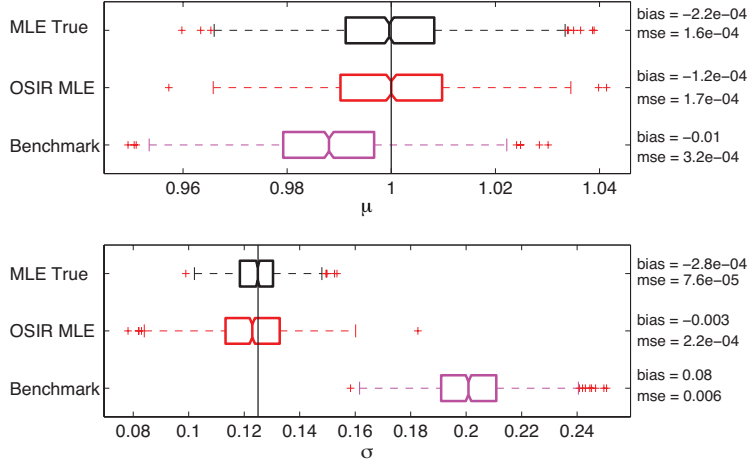}
\caption{Boxplot of the distributions of the parameters $\mu$ (top) and $\sigma$ (bottom) as estimated by maximum likelihood estimation using the ``true" event times (black), the OSIR filter-based estimates (red) and the benchmark estimates (magenta). Each boxplot shows the median and its 95$\%$ confidence intervals (notches). The boxes mark the 25th and 75th percentile (the interquartile range) of the empirical distribution. The whiskers mark 1.5 times the interquartile range and the red pluses are outliers beyond this range. 
 }
\label{fig:boxplot}
\end{figure}

We created 1000 replicas of a 
100-event point process
with the usual parameters and estimated the parameters of each catalog
using the OSIR particle filter,
the benchmark and the ``true" event times. Because of the stochastic
nature of the estimation problem, this resulted in a distribution of
parameters for each method. Figure \ref{fig:boxplot} compares the
distributions of the parameter estimates resulting from each of the
three methods by summarizing information in a boxplot. The main
observation from Figure \ref{fig:boxplot} is the lack of agreement
between the theoretical values of the parameters and the benchmark
medians. The benchmark, on average, underestimates $\mu$ by over 1$\%$
and overestimates $\sigma$ by a factor of almost 2. The benchmark estimate
of $\sigma$ is so poor that not a single estimate is close to the
``true" constant.  In contrast, the OSIR estimates
are much closer to the ``true" values than the
benchmark. While there is some evidence that $\sigma$ tends to be
slightly underestimated, the median estimate of $\mu$ is almost
indistinguishable from the ``true" value. Moreover, the distributions of
the parameter estimates are very similar to the distributions of the
parameters based on the ``true" event times. 

Two measures of the quality of an estimator are the mean-square error
and the bias. The former is the average squared difference between the
estimated value and the theoretical value, while the latter is the
difference between the mean estimate and the ``true" value. Figure
\ref{fig:boxplot} also shows the bias and mean-square errors for each method. The bias of the
 benchmark method is by far the greatest for the parameter $\sigma$. In
 contrast, the bias and the mean-square error of the OSIR particle filter 
are much closer to the values of the estimates based on the exact event times. 

This implies that the OSIR 
estimates come close to recovering the statistical properties of the
maximum likelihood estimator using the exact event times. 
The OSIR particle filter 
does not have exact event times to estimate the parameters and this loss
of information is necessarily reflected in the slightly broader
distributions (more uncertainty) and the slight bias in $\sigma$. At the
same time, the OSIR particle filter
obtains much better parameter estimates using only the observed event times. Figure \ref{fig:boxplot} shows that these estimates greatly improve on the biased benchmark estimates both in consistency and in variance. 

We also tested how the OSIR particle filter compares to the benchmark in terms of log-likelihood scores when the parameters are being estimated from an observed (but synthetic) data set, rather than using the exact parameters as in section \ref{subsec:Performance}. Using again the 1,000 replicas of a 100-event point process from which we estimated maximum (marginal) likelihood parameters, we used the maximum likelihood estimates of both the OSIR particle filter and the benchmark to calculate the cumulative log-likelihood ratios. We found that in $98.3\%$ of the 1,000 replicas, the log-likelihood score of the OSIR particle filter was larger than that of the benchmark. Only in 1.7$\%$ of all cases did the benchmark result in higher likelihood scores than the OSIR particle filter. Given the stochastic nature of the problem, one expects an overlap of
the two likelihood distributions. However, the OSIR
estimates, on average, provide a significantly higher likelihood score, along with better parameter
estimates. This concludes our proof-of-concept that
data assimilation implemented using sequential Monte Carlo methods can be successfully used as a technique for earthquake forecasting using renewal point-processes.

\section{Conclusion}
\label{sec:Con}

In this article, we have shown the potential benefits and the
feasibility of data assimilation-based earthquake forecasting for a
simple renewal process observed in noise. We used sequential Monte Carlo
methods, a flexible set of simulation-based methods for sampling from
arbitrary distributions, to represent the posterior distributions of the
exact event times given noisy observed event times. We showed that a
particular particle filter, 
the Optimal Sampling Importance Resampling (OSIR) filter
that uses the optimal importance density for sampling the posterior and that includes a resampling step to rejuvenate the particles, can solve this particular pedagogical example for an arbitrary number of events and may thus be useful for realistic problems. 

The forecasts based on the particle filter are significantly better than those of a benchmark forecast that ignores potential errors in the observed event times. We measured the improvement of our data assimilation-based method over the uncertainty-ignoring benchmark method by likelihood-based scores. In particular, we used the marginal complete data likelihood, the denominator in Bayes' theorem, to judge the quality of a forecast in the presence of observational noise. Parameters of the renewal process can be estimated by maximizing the marginal complete likelihood function. The particle filter parameter estimates were shown to be significantly less biased than those of the benchmark method. The marginal likelihood may further be used for hypothesis tests and model inference and comparison. 

This pedagogical example can be generalized to more realistic settings relevant to real earthquake forecasts and their evaluation, such as in RELM and CSEP. Data assimilation provides a powerful vehicle for earthquake forecasting and validation in the presence of the known data uncertainties. Another area of application lies in  model inference from paleoseismic data sets with complex conditional likelihood, which may benefit greatly from sequential Monte Carlo methods for Bayesian estimation. For point-process models of clustered seismicity, the relevant data uncertainties are probably to be found in the magnitude and location measurements rather than the occurrence times. It may also be interesting to extend the data assimilation approach to models of Coulomb stress changes, that are particularly influenced by location errors. There are a wide array of models that might benefit from data assimilation. We hope this article stimulates some interest in this area.

\section*{Acknowledgments}

This study was partially supported by the Swiss ETH CCES project
 EXTREMES (MJW and DS) and
Office of Naval Research grants N00014040191 and N000140910418 (KI). 


\end{document}